\documentclass{emulateapj}

\usepackage{amsmath}
\usepackage{natbib}
\usepackage{graphicx}
\usepackage{epsf}
\usepackage{color}
\usepackage{threeparttable}
\usepackage{comment}
\usepackage{epsfig}
\usepackage{xspace}
\usepackage[usenames,dvipsnames,svgnames]{xcolor}
\DeclareGraphicsExtensions{.jpg,.pdf,.png,.eps,.ps}

 \newcommand{\LCDM}{\mbox{$\Lambda$CDM}\xspace}
 \newcommand{\LCDMnospace}{\mbox{$\Lambda$CDM}}
 \newcommand{\omk}{\mbox{$\Omega_{k}$}}

 \newcommand{\ltsima}{$\; \buildrel < \over \sim \;$}
 \newcommand{\ltsim}{\lower.5ex\hbox{\ltsima}}

 \newcommand{\five}{$5^\mathrm{h} 30^\mathrm{m}$}

 \newcommand{\sqdeg}{\ensuremath{\mathrm{deg}^2}}

 \newcommand{\neff}{\ensuremath{N_\mathrm{eff}}\xspace}

 \newcommand{\deltaR}{\ensuremath{\Delta_R^2}}
 \newcommand{\nrun}{\ensuremath{dn_s/d\ln k}\xspace}
 \newcommand{\alens}{\ensuremath{A_{L}}}
 \newcommand{\ns}{\ensuremath{n_{s}}}
 \newcommand{\ho}{\ensuremath{H_{0}}\xspace}
 
 \newcommand{\sumnu}{\ensuremath{\Sigma m_\nu}\xspace} 
 \newcommand{\wmap}{\textit{WMAP}\xspace} 
 \newcommand{\wseven}{\textit{WMAP}7} 
 
 \newcommand{\nsCmb}{\ensuremath{0.9623\pm0.0097}}
 \newcommand{\nsCmbHo}{\ensuremath{0.9638\pm0.0090}}
 \newcommand{\nsCmbBao}{\ensuremath{0.9515\pm0.0082}}
 \newcommand{\nsCmbHoBao}{\ensuremath{0.9538\pm0.0081}}

 \newcommand{\nsCdfCmb}{\ensuremath{3.9}}
 \newcommand{\nsCdfCmbHo}{\ensuremath{4.0}}
 \newcommand{\nsCdfCmbBao}{\ensuremath{6.0}}
 \newcommand{\nsCdfCmbHoBao}{\ensuremath{5.7}}
 \newcommand{\nsCdfCmbHoBaoNeff}{\ensuremath{2.5}}

 \newcommand{\nsProbCmb}{\ensuremath{4 \times 10^{-5}}}
 \newcommand{\nsProbCmbHo}{\ensuremath{3.1 \times 10^{-5}}}
 \newcommand{\nsProbCmbBao}{\ensuremath{1.1 \times 10^{-9}}}
 
 \newcommand{\nsProbCmbHoBaoNeff}{\ensuremath{6.1 \times 10^{-3}}}

 \newcommand{\alensCdfSpt}{\ensuremath{5.9}}
 \newcommand{\alensProbSpt}{\ensuremath{1.3 \times 10^{-9}}}
 \newcommand{\alensCdfCmb}{\ensuremath{8.1}}
 \newcommand{\alensProbCmb}{\ensuremath{2.4 \times 10^{-16}}}
 \newcommand{\alensCmb}{\ensuremath{0.86^{+0.15 (+0.30)}_{-0.13 (-0.25)}}}

 \newcommand{\omkCmb}{\ensuremath{-0.003^{+0.014}_{-0.018}}}
 \newcommand{\omkCmbHo}{\ensuremath{0.0018\pm0.0048}}
 \newcommand{\omkCmbBao}{\ensuremath{-0.0089\pm0.0043}}
 \newcommand{\omkCmbHoBao}{\ensuremath{-0.0059\pm0.0040}}
 \newcommand{\omlCdfCmb}{\ensuremath{5.4}}

 \def\microKsq{\mu{\mbox{K}}^2}
 \def \zero {\textsc{ra5h30dec-55}}
 \def \one {\textsc{ra23h30dec-55}}
 \def \three {\textsc{ra21hdec-60}}
 \def \four {\textsc{ra3h30dec-60}}
 \def \five {\textsc{ra21hdec-50}}
 \def \six {\textsc{ra4h10dec-50}}
 \def \seven {\textsc{ra0h50dec-50}}
 \def \eight {\textsc{ra2h30dec-50}}
 \def \nine {\textsc{ra1hdec-60}}
 \def \ten {\textsc{ra5h30dec-45}}
 \def \eleven {\textsc{ra6h30dec-55}}
 \def \twelve {\textsc{ra23hdec-62.5}}
 \def \thirteen {\textsc{ra21hdec-42.5}}
 \def \fourteen {\textsc{ra22h30dec-55}}
 \def \fifteen {\textsc{ra23hdec-45}}
 \def \sixteen {\textsc{ra6hdec-62.5}}
 \def \seventeen {\textsc{ra3h30dec-42.5}}
 \def \eighteen {\textsc{ra1hdec-42.5}}
 \def \nineteen {\textsc{ra6h30dec-45}}

 \def\cl{$C_{\ell}$\xspace}
 \def\clnospace{\ensuremath{C_{\ell}}}
 \def\dl{$D_{\ell}$}

\def\KICPChicago{1}
\def\PhysicsUChicago{2}
\def\Berkeley{3}
\def\Davis{4}
\def\UChicago{5}
\def\EFIChicago{6}
\def\AAUChicago{7}
\def\Argonne{8}
\def\NIST{9}
\def\McGill{10}
\def\Colorado{11}
\def\NASA{12}
\def\LBNL{13}
\def\Caltech{14}
\def\Michigan{15}
\def\Munich{16}
\def\ExcellenceCluster{17}
\def\MPE{18}
\def\CaseWestern{19}
\def\Minnesota{20}
\def\ArtInstChicago{21}
\def\CfA{22}
\def\Dunlap{23}
\def\UToronto{24}
\def\BCCP{25}

\begin{document}
\title{A Measurement of the Cosmic Microwave Background Damping Tail from the 2500-square-degree SPT-SZ survey}

\slugcomment{Submitted to \apj}

 \author{
  K.~T.~Story,\altaffilmark{\KICPChicago,\PhysicsUChicago}
  C.~L.~Reichardt,\altaffilmark{\Berkeley}
  Z.~Hou,\altaffilmark{\Davis}
  R.~Keisler,\altaffilmark{\KICPChicago,\PhysicsUChicago}
  K.~A.~Aird,\altaffilmark{\UChicago}
  B.~A.~Benson,\altaffilmark{\KICPChicago,\EFIChicago}
  L.~E.~Bleem,\altaffilmark{\KICPChicago,\PhysicsUChicago}
  J.~E.~Carlstrom,\altaffilmark{\KICPChicago,\PhysicsUChicago,\EFIChicago,\AAUChicago,\Argonne}
  C.~L.~Chang,\altaffilmark{\KICPChicago,\EFIChicago,\Argonne}
  H-M.~Cho,\altaffilmark{\NIST}
  T.~M.~Crawford,\altaffilmark{\KICPChicago,\AAUChicago}
  A.~T.~Crites,\altaffilmark{\KICPChicago,\AAUChicago}
  T.~de~Haan,\altaffilmark{\McGill}
  M.~A.~Dobbs,\altaffilmark{\McGill}
  J.~Dudley,\altaffilmark{\McGill}
  B.~Follin,\altaffilmark{\Davis}
  E.~M.~George,\altaffilmark{\Berkeley}
  N.~W.~Halverson,\altaffilmark{\Colorado}
  G.~P.~Holder,\altaffilmark{\McGill}
  W.~L.~Holzapfel,\altaffilmark{\Berkeley}
  S.~Hoover,\altaffilmark{\KICPChicago,\PhysicsUChicago}
  J.~D.~Hrubes,\altaffilmark{\UChicago}
  M.~Joy,\altaffilmark{\NASA}
  L.~Knox,\altaffilmark{\Davis}
  A.~T.~Lee,\altaffilmark{\Berkeley,\LBNL}
  E.~M.~Leitch,\altaffilmark{\KICPChicago,\AAUChicago}
  M.~Lueker,\altaffilmark{\Caltech}
  D.~Luong-Van,\altaffilmark{\UChicago}
  J.~J.~McMahon,\altaffilmark{\Michigan}
  J.~Mehl,\altaffilmark{\Argonne,\KICPChicago}
  S.~S.~Meyer,\altaffilmark{\KICPChicago,\PhysicsUChicago,\EFIChicago,\AAUChicago}
  M.~Millea,\altaffilmark{\Davis}
  J.~J.~Mohr,\altaffilmark{\Munich,\ExcellenceCluster,\MPE}
  T.~E.~Montroy,\altaffilmark{\CaseWestern}
  S.~Padin,\altaffilmark{\KICPChicago,\AAUChicago,\Caltech}
  T.~Plagge,\altaffilmark{\KICPChicago,\AAUChicago}
  C.~Pryke,\altaffilmark{\Minnesota}
  J.~E.~Ruhl,\altaffilmark{\CaseWestern}
  J.T.~Sayre\altaffilmark{\CaseWestern}
  K.~K.~Schaffer,\altaffilmark{\KICPChicago,\EFIChicago,\ArtInstChicago}
  L.~Shaw,\altaffilmark{\McGill}
  E.~Shirokoff,\altaffilmark{\Berkeley} 
  H.~G.~Spieler,\altaffilmark{\LBNL}
  Z.~Staniszewski,\altaffilmark{\CaseWestern}
  A.~A.~Stark,\altaffilmark{\CfA}
  A.~van~Engelen,\altaffilmark{\McGill}
  K.~Vanderlinde,\altaffilmark{\Dunlap,\UToronto}
  J.~D.~Vieira,\altaffilmark{\Caltech}
  R.~Williamson,\altaffilmark{\KICPChicago,\AAUChicago} and
  O.~Zahn\altaffilmark{\BCCP}
 }

\altaffiltext{\KICPChicago}{Kavli Institute for Cosmological Physics, University of Chicago, 5640 South Ellis Avenue, Chicago, IL, USA 60637}
\altaffiltext{\PhysicsUChicago}{Department of Physics, University of Chicago, 5640 South Ellis Avenue, Chicago, IL, USA 60637}
\altaffiltext{\Berkeley}{Department of Physics, University of California, Berkeley, CA, USA 94720}
\altaffiltext{\Davis}{Department of Physics, University of California, One Shields Avenue, Davis, CA, USA 95616}
\altaffiltext{\UChicago}{University of Chicago, 5640 South Ellis Avenue, Chicago, IL, USA 60637}
\altaffiltext{\EFIChicago}{Enrico Fermi Institute, University of Chicago, 5640 South Ellis Avenue, Chicago, IL, USA 60637}
\altaffiltext{\AAUChicago}{Department of Astronomy and Astrophysics, University of Chicago, 5640 South Ellis Avenue, Chicago, IL, USA 60637}
\altaffiltext{\Argonne}{Argonne National Laboratory, 9700 S. Cass Avenue, Argonne, IL, USA 60439}
\altaffiltext{\NIST}{NIST Quantum Devices Group, 325 Broadway Mailcode 817.03, Boulder, CO, USA 80305}
\altaffiltext{\McGill}{Department of Physics, McGill University, 3600 Rue University, Montreal, Quebec H3A 2T8, Canada}
\altaffiltext{\Colorado}{Department of Astrophysical and Planetary Sciences and Department of Physics, University of Colorado, Boulder, CO, USA 80309}
\altaffiltext{\NASA}{Department of Space Science, VP62,NASA Marshall Space Flight Center,Huntsville, AL, USA 35812}
\altaffiltext{\LBNL}{Physics Division, Lawrence Berkeley National Laboratory, Berkeley, CA, USA 94720}
\altaffiltext{\Caltech}{California Institute of Technology, MS 249-17, 1216 E. California Blvd., Pasadena, CA, USA 91125}
\altaffiltext{\Michigan}{Department of Physics, University of Michigan, 450 Church Street, Ann  Arbor, MI, USA 48109}
\altaffiltext{\Munich}{Department of Physics, Ludwig-Maximilians-Universit\"{a}t,Scheinerstr.\ 1, 81679 M\"{u}nchen, Germany}
\altaffiltext{\ExcellenceCluster}{Excellence Cluster Universe, Boltzmannstr.\ 2, 85748 Garching, Germany}
\altaffiltext{\MPE}{Max-Planck-Institut f\"{u}r extraterrestrische Physik,Giessenbachstr.\ 85748 Garching, Germany}
\altaffiltext{\CaseWestern}{Physics Department, Center for Education and Research in Cosmology and Astrophysics, Case Western Reserve University,Cleveland, OH, USA 44106}
\altaffiltext{\Minnesota}{Department of Physics, University of Minnesota, 116 Church Street S.E. Minneapolis, MN, USA 55455}
\altaffiltext{\ArtInstChicago}{Liberal Arts Department, School of the Art Institute of Chicago, 112 S Michigan Ave, Chicago, IL, USA 60603}
\altaffiltext{\CfA}{Harvard-Smithsonian Center for Astrophysics, 60 Garden Street, Cambridge, MA, USA 02138}
\altaffiltext{\Dunlap}{Dunlap Institute for Astronomy \& Astrophysics, University of Toronto, 50 St George St, Toronto, ON, M5S 3H4, Canada}
\altaffiltext{\UToronto}{Department of Astronomy \& Astrophysics, University of Toronto, 50 St George St, Toronto, ON, M5S 3H4, Canada}
\altaffiltext{\BCCP}{Berkeley Center for Cosmological Physics, Department of Physics, University of California, and Lawrence Berkeley National Laboratory, Berkeley, CA, USA 94720}

\email{kstory@uchicago.edu}

\begin{abstract}
We present a measurement of the cosmic microwave background (CMB) temperature power spectrum using data from the recently completed South Pole Telescope Sunyaev-Zel'dovich (SPT-SZ) survey.
 This measurement is made from observations of 2540 deg$^2$ of sky with arcminute resolution at $150\,$GHz, and improves upon previous measurements using the SPT by tripling the sky area. 
 We report CMB temperature anisotropy power over the multipole range $650<\ell<3000$.
 We fit the SPT bandpowers, combined with the seven-year Wilkinson Microwave Anisotropy Probe (\wseven) data, with a six-parameter \LCDM cosmological model and find that the two datasets are consistent and well fit by the model.
 Adding SPT measurements significantly improves \LCDM parameter constraints; in particular, the constraint on $\theta_s$ tightens by a factor of 2.7.
 The impact of gravitational lensing is detected at $\alensCdfCmb\,\sigma$, the most significant detection to date.
 This sensitivity of the SPT+\wseven{} data to lensing by large-scale structure at low redshifts allows us to constrain the mean curvature of the observable universe with CMB data alone to be $\omk=\omkCmb$.
 Using the SPT+\wseven{} data, we measure the spectral index of scalar fluctuations to be $n_s=\nsCmb$ in the \LCDM model, a $\nsCdfCmb\,\sigma$ preference for a scale-dependent spectrum with $\ns<1$.  
 The SPT measurement of the CMB damping tail helps break the degeneracy that exists between the tensor-to-scalar ratio $r$ and \ns{} in large-scale CMB measurements, leading to an upper limit of $r<0.18$ (95\%\,C.L.) in the \LCDMnospace+$r$ model.
 Adding low-redshift measurements of the Hubble constant ($H_0$) and the baryon acoustic oscillation (BAO) feature to the SPT+\wseven{} data leads to further improvements. 
The combination of SPT+\wseven+$H_0$+BAO constrains $n_s=\nsCmbHoBao$ in the \LCDM model, a $\nsCdfCmbHoBao\,\sigma$ detection of $\ns < 1$, and places an upper limit of $r<0.11$ (95\%\,C.L.) in the \LCDMnospace+$r$ model.
 These new constraints on \ns{} and $r$ have significant implications for our understanding of inflation, which we discuss in the context of selected single-field inflation models.

\end{abstract}


\bigskip\bigskip


\section{Introduction}
\label{sec:intro}

Over the past two decades, measurements of the cosmic microwave background (CMB) have provided profound insight into the nature of the universe.  
 Detailed information about the composition and evolution of the universe is encoded in the temperature and polarization anisotropy of the CMB.
 Measuring this anisotropy enables powerful tests of cosmological theory.
 On degree scales, CMB anisotropy is primarily generated by the acoustic oscillations of the primordial plasma in the early universe.
 The Wilkinson Microwave Anisotropy Probe (\wmap) satellite has been used to measure these acoustic oscillations with cosmic variance-limited precision on angular scales corresponding to $\ell \lesssim 500$ \citep[hereafter \wseven{}]{komatsu11}. 
 On much smaller angular scales, primary CMB anisotropy becomes dominated by effects imprinted on the CMB at low redshift (so-called secondary anisotropy) and foregrounds; at millimeter wavelengths, this transition occurs at $\ell \sim 3000$.
 This small-angular scale millimeter-wavelength anisotropy has been measured by the South Pole Telescope \citep[SPT,][]{lueker10, shirokoff11, reichardt12b} and the Atacama Cosmology Telescope \citep[ACT,][]{fowler10, das11b}.

The anisotropy in the CMB at intermediate angular scales, $1000 \lesssim \ell \lesssim 3000$, is often referred to as the ``damping tail'' since the anisotropy power on these angular scales is damped by photon diffusion during recombination \citep{silk68}. 
 Adding measurements of the damping tail to large-scale CMB measurements considerably tightens the resulting cosmological constraints. 
 Therefore measurement on smaller angular scales are sensitive to the photon diffusion scale during recombination.
 The wider range of angular scales also enables better constraints on the sound horizon at recombination (by measuring more acoustic peaks) and the slope of the primordial power spectrum. 
 Finally, although tensor perturbations from cosmic inflation add CMB power only at very large angular scales, the effect of these tensor perturbations is 
degenerate with changes in \ns{} in large-scale measurements.  
 Damping tail measurements help break this degeneracy, thus tightening constraints on the level of tensor perturbations.

In the past few years, there have been several increasingly precise measurements of the CMB damping tail, including the Arcminute Cosmology Bolometer Array Receiver \citep[ACBAR,][]{reichardt09a}, QUaD \citep{brown09, friedman09},  ACT \citep{das11}, and SPT \citep[hereafter K11]{keisler11}.  
 The most precise published measurement of the CMB damping tail prior to this work comes from the first 790 deg$^2$ of the South Pole Telescope Sunyaev-Zel'dovich (SPT-SZ) survey (K11).

In this paper, we present a measurement of the power spectrum from the third acoustic peak through the CMB damping tail, covering the range of angular scales corresponding to multipoles $650 < \ell < 3000$.
 This power spectrum is calculated from the complete SPT-SZ survey covering $2540\, {\rm deg}^2$ of sky, and improves upon the results presented in K11 by expanding the sky coverage by a factor of three.   

We present constraints from this measurement on the standard \LCDM{} model of cosmology, then extend the model to quantify the amplitude of gravitational lensing of the CMB. 
 We use this sensitivity to gravitational lensing by large-scale structure at low redshifts to measure the mean curvature of the observable universe from CMB data alone.
 We also consider models including tensor perturbations, and explore implications of the resulting parameter constraints for simple models of inflation. 
 Adding low-redshift measurements of the Hubble constant ($H_0$) and the baryon acoustic oscillation (BAO) feature to the CMB data further tightens parameter constraints, and we present combined parameter constraints for each of the above model extensions. 
 The implications of the SPT power spectrum for a larger range of extensions to the standard cosmological model are explored more fully in a companion paper, \citet[hereafter H12]{hou12}. 

This paper is organized as follows.  
 We describe the SPT observations and data reduction in \S \ref{sec:data}. 
 We present the power spectrum calculation in \S \ref{sec:ps}.  
 We discuss tests for systematic errors in \S \ref{sec:nulltests}.
 We present the power spectrum measurement in \S \ref{sec:bandpowers}.
 In \S \ref{sec:constraints}, we outline our cosmological parameter fitting framework and present the resulting parameter constraints, then use these constraints to explore the implications for simple models of inflation.
 Finally, we conclude in \S \ref{sec:conclusion}.

\begin{figure*}
\begin{center}
    \includegraphics[width=0.85\textwidth]{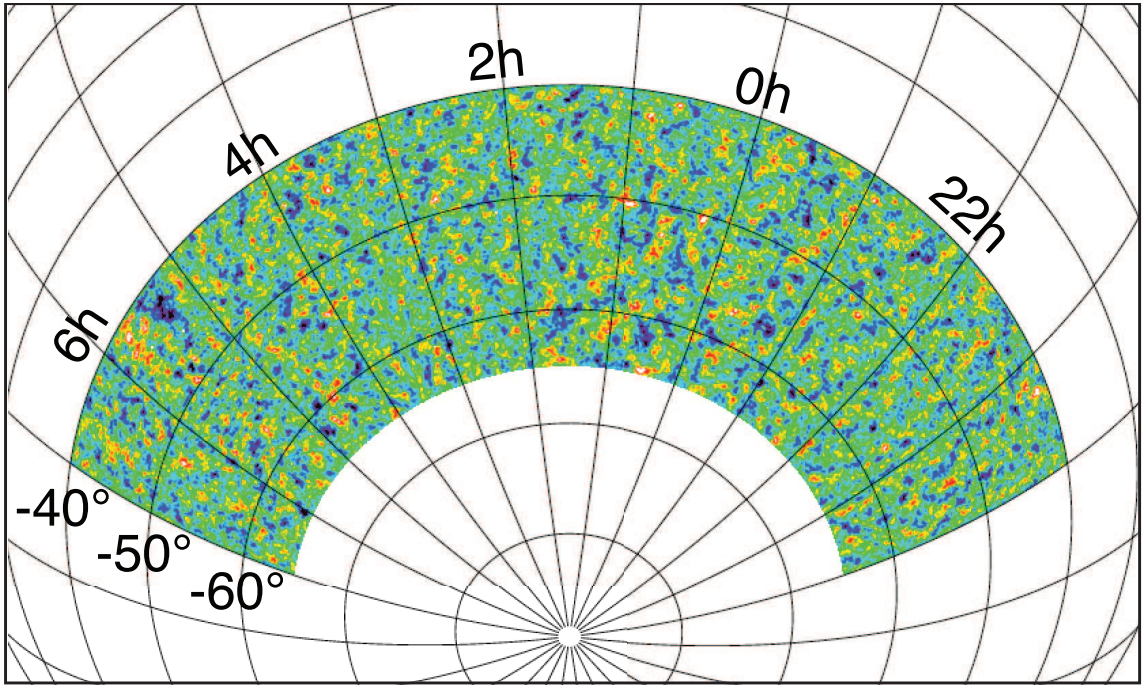}
\end{center}
\caption{The 2500 deg$^2$ SPT-SZ survey.  
We show the full survey region with lightly filtered 95 GHz data from the SPT, using the data and filters which best capture the degree-scale anisotropy of the CMB visible in this figure.
The power spectrum measurement reported in this paper is calculated from 2540 deg$^2$ of sky and analyzes 150 GHz data with a different high-pass filter, as described in \S~\ref{sec:mapmaking}.
} 
\label{fig:icon}
\end{figure*}


\section{Observations and Data Reduction}
\label{sec:data}

The SPT is a 10-meter diameter telescope located at the Amundsen-Scott South Pole station in Antarctica.  
 The first survey with the SPT, referred to as the ``SPT-SZ'' survey, was completed in November 2011 and covered a $\sim 2500$ deg$^2$ region of sky between declinations of -40$^\circ$ and -65$^\circ$ and right ascensions of 20h and 7h.
 The SPT-SZ survey is shown in Figure~\ref{fig:icon}.
 Here we present the first power spectrum measurement that uses data from the complete SPT-SZ survey.
 We use data from 2540 deg$^2$ of sky in this analysis.

This work uses observations and data reduction methods that are very similar to those described in K11. 
 In this section, we give an overview of the observations and data reduction, highlighting the differences with the treatment in K11, to which we refer the reader for a detailed treatment of the analysis methods.

\subsection{Observing Strategy and Fields}
\label{sec:fields_and_obs}

\begin{figure}
\begin{center}
    \includegraphics[width=0.48\textwidth]{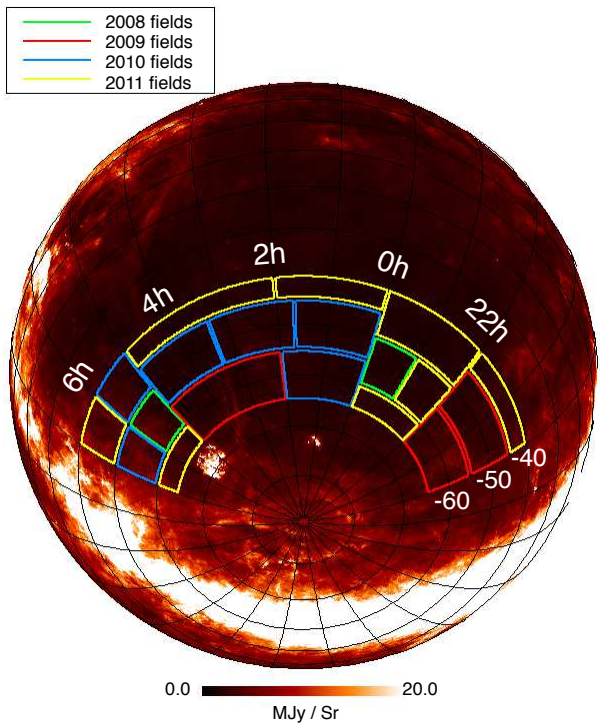}
\end{center}
\caption{The SPT was used to observe 2500 deg$^2$ over 19 individual fields, which are overlaid here on an orthographic projection of the IRAS $100 \, \mu$m dust map from \citet{schlegel98}.
 These observation fields were chosen to lie in regions of low dust emission (dark red).}
\label{fig:fields}
\end{figure}

From 2008 - 2011, the SPT was used to observe a contiguous $\sim2500$ deg$^2$ patch of sky to a noise level of approximately 18\,$\mu{\rm K}$-arcmin\footnote{Throughout this work, the unit K refers to equivalent fluctuations in the CMB temperature, i.e., the temperature fluctuation of a 2.73 K blackbody that would be required to produce the same power fluctuation.  The conversion factor is given by the derivative of the blackbody spectrum $\frac{dB_{\nu}}{dT}$, evaluated at 2.73 K.} 
 at 150\, GHz.\footnote{The SPT-SZ survey also includes data at 95 and 220\,GHz. 
However, this work uses only 150 GHz data since this observing band is the most sensitive for the SPT and the data from one observing band are sufficient to make high signal-to-noise maps of the CMB anisotropy.}
 This area of sky was observed in 19 contiguous sub-regions which we refer to as observation ``fields''.  
 In the basic survey strategy, the SPT was used to observe a single field until the desired noise level was reached before moving on to the next field. 
 Two fields were observed in 2008, three in 2009, five in 2010, and nine in 2011. 
 All nine fields from 2011 were observed to partial depth in 2010 in order to search for massive galaxy clusters, then re-observed in 2011 to achieve nominal noise levels.
 The results of that bright cluster search were published in \cite{williamson11}.
 In terms of sky area, this equates to observing 167\,\sqdeg{} in 2008, 574\,\sqdeg{} in 2009, 732\,\sqdeg{} in 2010, and 1067\,\sqdeg{} in 2011.
 The fields are shown in Figure~\ref{fig:fields}, and the field locations and sizes are presented in Table~\ref{tab:fields}.

Both fields from 2008 (\zero{} and \one) were re-observed in later years to achieve lower than normal noise levels.
 In this analysis, we use data from only one year for each field because the beam and noise properties vary slightly between years. 
 This choice simplifies the analysis without affecting the results as the bandpower uncertainties remain sample variance dominated (see \S \ref{sec:cov}).

The SPT is used to observe each field in the following manner.
 The telescope starts in one corner of the observation field, slews back and forth across the azimuth range of the field, and then executes a step in elevation, repeating this pattern until the entire field has been covered.
 This constitutes a single observation of the field, and takes from 30 minutes to a few hours, depending on the specific field being observed.
 Azimuthal scan speeds vary between fields, ranging from 0.25 to 0.42 degrees per second on the sky.
 The starting elevation positions of the telescope are dithered by between $0.3^{\prime}$ and $1.08^{\prime}$ to ensure uniform coverage of the region in the final coadded map. 

In four of the 2008 and 2009 fields, \one, \three, \four, and \five, observations were conducted with a ``lead-trail'' strategy.  
 In this observation strategy, the field is divided into two halves in right ascension.  
 The ``lead'' half is observed first, immediately followed by the ``trail'' half in a manner such that both halves are observed over the same azimuthal range. 
 If necessary, the lead-trail data could be analyzed in a way that cancels ground pickup. 
 In this analysis, we combine lead-trail pairs into single maps, and verify that contamination from ground pickup is negligible -- see below and \S~\ref{sec:nulltests} for details.

We apply several (often redundant) data quality cuts on individual observations using the following criteria: 
 map noise, noise-based bolometer weight, the product of median bolometer weight with map noise, and the sum of bolometer weights over the full map. 
 For these cuts, we remove outliers both above and below the median value for each field. 
 We do not use observations that are flagged by one or more of these cuts. 
 We also flag observations with only partial field coverage. 
 Finally, we cut maps that were made from observations in azimuth ranges that could be more susceptible to ground pickup over the angular scales of interest.  
 We use ``ground-centered'' maps to measure ground pickup on large ($\ell \sim50$) scales, and cut observations that were made at the azimuths with the worst 5\% ground pickup to minimize potential ground-pickup on smaller angular scales.  
 Although this cut does have an impact on our null tests (see \S~\ref{sec:nulltests}), we emphasize that it does not significantly change the power spectrum, the precision of which is limited by sample variance.

\begin{table*}[]
\centering
\caption[SPT Fields]{The fields observed with the SPT between 2008 and 2011} \small
\begin{tabular}{ | l  r r r r c | }
\hline
Name & R.A. ($^\circ$) & Decl. ($^\circ$) & $\Delta$R.A. ($^\circ$) & $\Delta$Decl. ($^\circ$) & Effective Area (deg$^2$)\\
\hline \hline
\zero      &  82.7 &   -55.0 &             15 &             10 &    84 \\
\one       & 352.5 &   -55.0 &             15 &             10 &    83 \\
\three     & 315.0 &   -60.0 &             30 &             10 &   155 \\
\four      &  52.5 &   -60.0 &             45 &             10 &   227 \\
\five      & 315.0 &   -50.0 &             30 &             10 &   192 \\
\six       &  62.5 &   -50.0 &             25 &             10 &   156 \\
\seven     &  12.5 &   -50.0 &             25 &             10 &   157 \\
\eight     &  37.5 &   -50.0 &             25 &             10 &   157 \\
\nine      &  15.0 &   -60.0 &             30 &             10 &   152 \\
\ten       &  82.5 &   -45.0 &             15 &             10 &   109 \\
\eleven    &  97.5 &   -55.0 &             15 &             10 &    85 \\
\twelve    & 345.0 &   -62.5 &             30 &              5 &    75 \\
\thirteen  & 315.0 &   -42.5 &             30 &              5 &   121 \\
\fourteen  & 337.5 &   -55.0 &             15 &             10 &    84 \\
\fifteen   & 345.0 &   -45.0 &             30 &             10 &   217 \\
\sixteen   &  90.0 &   -62.5 &             30 &              5 &    75 \\
\seventeen &  52.5 &   -42.5 &             45 &              5 &   179 \\
\eighteen  &  15.0 &   -42.5 &             30 &              5 &   119 \\
\nineteen  &  97.5 &   -45.0 &             15 &             10 &   111 \\
\hline
Total      & & & & &   2540\\
\hline
\end{tabular}
\label{tab:fields}
\tablecomments{The locations and sizes of the fields observed by the SPT between 2008 and 2011.  For each field we give the center of the field in Right Ascension (R.A.) and declination (decl.), the nominal extent of the field in Right Ascension and declination, and the effective field area as defined by the window (see \S \ref{sec:window}).}
\end{table*}

\subsection{Map-making: Time Ordered Data to Maps}
\label{sec:mapmaking}

As the SPT scans across the sky, the response of each detector is recorded as time-ordered data (TOD).  
 These TOD are recorded at 100 Hz and have a Nyquist frequency of 50 Hz, which corresponds to a multipole number parallel to the scan direction ($\ell_x$) between 72,000 and 43,000 at the SPT scan speeds.
 Since we only report the power spectrum up to $\ell = 3000$, we can benefit computationally by reducing the sampling rate.
 We choose a low-pass filter and down-sampling factor based on each field's scan speed such that they affect approximately the same angular scales. 
 We use a down-sampling factor of 6 for 2008 and 2009, and 4 for 2010 and 2011 with associated low-pass filter frequencies of 7.5 and 11.4 Hz  respectively.
 These filtering choices remove a negligible amount of power in the signal band.

Next, the down-sampled TOD are bandpass filtered between $\ell_x = 270$ and 6600. 
 The low-pass filter is necessary to avoid aliasing high-frequency noise to lower frequencies during map-making. 
 The high-pass filter reduces low-frequency noise from the atmosphere and instrumental readout. 
 The high-pass filter is implemented by fitting each bolometer's TOD (from a single azimuthal scan across the field) to a model consisting of low-frequency sines and cosines and a fifth-order polynomial.  
 The best-fit model is then subtracted from the TOD.  
 During the filtering, we mask regions of sky within 5 arcminutes of point sources with fluxes of S$_{150 \hbox{\tiny{GHZ}}} > 50$ mJy.
 These regions are also masked in the power spectrum analysis, see \S~\ref{sec:window}.
 
At this stage, the TOD retain signal from the atmosphere that is correlated between detectors.  
 We remove the correlated signal by subtracting the mean signal across each detector module for every time sample.\footnote{The SPT-SZ focal plane has a hexagonal geometry with six triangular bolometer modules, each with $\sim$160 detectors.  
 Each module is configured with a set of filters that determines its observing frequency of 95, 150, or 220 GHz.}
 This process acts as an approximately isotropic high-pass filter.

The filtered TOD are made into maps using the process described by K11.
 The data from each detector receive a weight based on the power spectral density of that detector's calibrated TOD in the 1-3 Hz band.
 This band corresponds approximately to the signal band of this analysis.  
We have calculated the level of bias introduced by using the full (signal + noise) power to calculate the detector weights, as opposed to using the noise power only \citep{dunner12}, and we find that the level of bias is completely negligible ($\le 0.01\%$ in power).
 The detector data are binned into maps with $1'$ pixels based on the telescope pointing information.  
In the power spectrum analysis presented in \S \ref{sec:ps}, we adopt the flat-sky approximation, where the wavenumber $k$ is equivalent to multipole moment $\ell$ and spherical harmonic transforms are replaced by Fourier transforms.  
 We project from the curved celestial sky to flat-sky maps with the oblique Lambert equal-area azimuthal projection \citep{snyder87}.

\subsection{Beam Functions}
\label{sec:beams}

A precise measurement of the SPT beam --- the optical response as a function of angle --- is needed to calibrate the angular power spectrum as a function of multipole.  
 We summarize the method used to measure the SPT beams and refer the reader to K11 or \cite{schaffer11} for a more detailed description.

The average 150 GHz beam is measured for each year using a combination of maps from Jupiter, Venus, and the 18 brightest point sources in the CMB fields.
 The maps of Jupiter are used to measure the beam outside a radius of $4^\prime$, while the maps of the bright point sources are used to measure the beam inside that radius.  
 Maps of Venus are used to join the inner and outer beam maps into a composite beam map.  
 The maps of the planets are not used to estimate the very inner beam due to a non-linear detector response when directly viewing bright sources and the non-negligible angular size of the planets compared to the beam.
 We use the composite beam map to measure the beam function $B_\ell$, defined as the azimuthally averaged Fourier transform of the beam map.  
 We consider uncertainties in the measurement of $B_\ell$ arising from several statistical and systematic effects, such as residual atmospheric noise in the maps of Venus and Jupiter, and account for known inter-year correlations of some of these sources of uncertainty.  
 The parameter constraints quoted in this work are not sensitive to the calculated beam uncertainties; we have tested increasing our beam uncertainties by a factor of two and have seen no significant impact on the resulting cosmological parameter fits.

A nearly identical beam treatment  was used by K11.  
 The main difference is that the beam function was normalized to unity at $\ell=350$ in K11 rather than $\ell=750$ in this work.  
 The average multipole of our calibration region is close to $\ell=750$, and this choice of normalization scale better decouples the beam and calibration uncertainties.

\subsection{Calibration}
\label{sec:calibration}

The observation-to-observation relative calibration of the TOD is determined from repeated measurements of a galactic H{\small \,II} region, RCW38. 
 As in K11 and \citet{reichardt12b}, the absolute calibration is determined by comparing the SPT and \wseven{} power in several $\ell$-bins over the multipole range $\ell \in [650,1000]$. 
 We use the same $\ell$-bins for both experiments: seven bins with $\delta\ell = 50$. 
 This calibration method is model-independent, requiring only that the CMB power in the SPT fields is statistically representative of the all-sky power. 
 We estimate the uncertainty in the SPT power calibration to be 2.6\%. 
 The calibration uncertainty is included in the covariance matrix; this treatement is equivalent to including an additional calibration parameter with a 2.6\% Gaussian uncertainty in the cosmological parameter fits.
 We have also cross-checked this method against a map-based calibration method, in which we calculate the cross-spectrum between identically filtered SPT and WMAP maps over 1250 deg$^2$ of sky, and find that the calibrations between these two methods are consistent, though the map-based calibration uncertainties are larger.
 We do not find the parameter constraints quoted in this work to be sensitive to the calibration uncertainty; changing the calibration uncertainty by a factor of two in either direction has no significant impact on the resulting cosmological parameter fits.

\section{Power Spectrum}
\label{sec:ps}

In this section, we describe the power spectrum calculation. 
 This analysis closely follows the analysis developed by \citet{lueker10} and used by K11; we refer the reader to those papers for a more detailed description.
 We refer to the average power over a given range of $\ell$ values as a \textit{bandpower}.
 We use a pseudo-\cl method under the flat-sky approximation as described in \S \ref{sec:mapmaking}.
 The power spectra are calculated independently in each of the 19 fields, then combined into the final result. 
 We report bandpowers in terms of \dl, which is defined as
\begin{equation}
  D_{\ell} = \frac{\ell (\ell + 1)}{2\pi} C_{\ell} \, .
\end{equation}
To calculate $D_{\ell}$, we use a cross-spectrum bandpower estimator as described in \S~\ref{sec:crossspectra}, which has the advantage of being free of noise bias; see \citet{lueker10} for a more detailed description.

\subsection{Maps}
\label{sec:maps}

The basic input to the cross-spectrum estimator is a set of maps for a given field, each with independent noise.
 For most fields, this input set is comprised of maps from single observations. 
 Each observation has statistically independent noise because observations are temporally separated by at least an hour and the TOD have been high-pass filtered at $\sim0.2$\,Hz. 
 As in K11, for the four fields observed with a lead-trail strategy, we construct the input map set by combining lead-trail pairs into single maps.
 The \one{} field was observed using comparatively large elevation steps and hence has less uniform coverage. 
 For this field, the single maps that are the basic input to the cross-spectrum estimator are formed by combining two pairs of lead-trail observations. 
 Each pair is chosen to have different elevation dithers, leading to a more homogeneous field coverage.

\subsection{Window}
\label{sec:window}

For a given field, each of the maps is multiplied by the same window \textbf{W} in order to avoid sharp edges at map boundaries, control overlap between adjacent fields, and remove bright point sources.
 Each window is the product of an apodization mask with a point source mask.  
 The apodization masks are calculated by applying a $1^\circ$ taper using a Hann function to the edges of the uniform coverage region of each field.
 The observations were designed such that the uniform coverage region overlaps between neighboring fields.
 We define our apodization windows such that the overlap region between adjacent fields contains a combined weight that approaches but never exceeds unity (the weight at the center of the field).  
 This process results in apodization windows that include marginally smaller regions of sky and fall off more slowly than the windows used in K11, which did not need to account for field overlap.

As was done in K11, we identify all point sources with 150 GHz flux $>$ 50 mJy.  
 Each of these point sources is masked with a $5'$-radius disk that is tapered outside the disk using a Gaussian taper with a width of $\sigma_{\hbox{taper}}=5'$.
 Point source masks remove $1.4\%$ of the total sky area. 
 Using previous measurements of the mm-wave point source population \citep{vieira10, shirokoff11}, we estimate that the power from residual point sources below this flux cut is $C_\ell \sim 1.3 \times 10^{-5} \microKsq$, or $D_\ell \sim 18~\microKsq \left(\frac{\ell}{3000}\right)^2$.  
 This power is approximately half the CMB anisotropy power at $\ell=3000$, the upper edge of the multipole range reported in this analysis.
 Further discussion of the point source model is reserved for \S~\ref{sec:model}.

\subsection{Cross-Spectra}
\label{sec:crossspectra}
The next step in calculating the power spectrum is to cross-correlate single maps from different observations of the same field.
 Each map is multiplied by the window for its field, zero-padded to the same size for all fields, then the Fourier transform of the map $\tilde{m}^{A}$ is calculated, where $A$ is the observation index.  
The resulting Fourier-space maps have pixels of size $\delta_{\ell}=5$ on a side.
We calculate the average cross spectrum between the maps of two observations $A$ and $B$ within an $\ell$-bin $b$:
\begin{equation}
\label{eqn:ddef}
 \widehat{D}^{AB}_b\equiv \left< \frac{\ell(\ell+1)}{2\pi}H_{\pmb{\ell}}Re[\tilde{m}^{A}_{\pmb{\ell}}\tilde{m}^{B*}_{\pmb{\ell}}] \right>_{\ell \in b}, 
\end{equation}
where $H_{\pmb{\ell}}$ is a two-dimensional weight array described below, and $\pmb{\ell}$ is a vector in two-dimensional $\ell$-space.  
Each field typically has about 200 single maps in the input set (see \S~\ref{sec:maps}), resulting in $\sim20,000$ cross-spectra.  
We average all cross-spectra $\widehat{D}^{AB}_b$ for $A \neq B$ to calculate a binned power spectrum $\widehat{D}_b$ for each field.  

Due to our observation strategy, the maps have statistically anisotropic noise; at fixed $\ell$, modes that oscillate perpendicular to the scan direction ($\ell_x = 0$) are noisier than modes that oscillate parallel to the scan direction.  
 This anisotropic noise -- and the filtering we apply to reduce the noise (see \S~\ref{sec:mapmaking}) -- causes different modes in a given $\ell$ bin to have different noise properties.
 As in K11, we use a two-dimensional weight $H_{\pmb{\ell}}$ which accounts for the anisotropic noise in the maps.
 We define the weight array according to
\begin{equation}
\label{eqn:2dweight}
H_{\pmb{\ell}} \propto (C_\ell^{\rm{th}} + N_{\pmb{\ell}})^{-2} \,,
\end{equation}
 where $C_\ell^{\rm{th}}$ is the theoretical power spectrum used in simulations described in \S \ref{sec:transfer}, and $N_{\pmb{\ell}}$ is the two-dimensional calibrated, beam-deconvolved noise power, which is calculated from difference maps in which the right-going scans are subtracted from the left-going scans.  
 The weight array is then smoothed with a Gaussian kernel of width $\sigma_{\ell}=450$ to reduce the scatter in the noise power estimate, and normalized to the maximum value in each annulus.  
$H_{\pmb{\ell}}$ is calculated independently for each observation field.

\subsection{Unbiased Spectra}
\label{sec:UnbiasedSpectra}

The power $\widehat{D}_b$ is a biased estimate of the true sky power, $D_b$, due to effects such as TOD filtering, projection effects, and mode mixing from the window.
 The biased and unbiased estimates are related by
\begin{equation}
\widehat{D}_{b^\prime} \equiv K_{b^\prime b} D_b \, ,
\end{equation} 
 where the $K$ matrix accounts for the effects of the beams, TOD filtering, pixelization, windowing, and band-averaging. 
 $K$ can be expanded as
\begin{equation}
\label{eqn:kdef}
K_{bb^\prime}=P_{b\ell}\left(M_{\ell\ell^\prime}[\textbf{W}]\,F_{\ell^\prime}B^{2}_{\ell^\prime}\right)Q_{\ell^\prime b^\prime}. 
\end{equation}
$Q_{\ell^\prime b^\prime}$ is the binning operator and $P_{b\ell}$ is its reciprocal \citep{hivon02}.  
The ``mode-coupling matrix'' $M_{\ell\ell^\prime}[\textbf{W}]$ accounts for mixing modes between multipole moments which arises from observing a finite portion of the sky.  
We calculate $M_{\ell\ell^\prime}[\textbf{W}]$ analytically from the window function $\textbf{W}$ following the prescription described by \cite{lueker10}.  
Over the range of multipoles reported in this analysis, the elements of the mode-coupling matrix depend only on the distance from the diagonal.  
$F_{\ell}$ is the transfer function due to TOD filtering and map pixelization, which is described in \S \ref{sec:transfer}.  
$B^{2}_{\ell}$ is the beam function described in \S \ref{sec:beams}.
For sufficiently large $\ell$-bins, the $K$ matrix is invertible, allowing an unbiased estimate of the true sky power:
\begin{equation}
D_b\equiv \left(K^{-1}\right)_{bb^\prime}\widehat{D}_{b^\prime} \, .
\end{equation}

\subsubsection{Simulations and the Transfer Function}
\label{sec:transfer}

The transfer function $F_{\ell}$ is calculated from end-to-end simulations.
 One hundred full-sky realizations are generated at a Healpix\footnote{http://healpix.jpl.nasa.gov} resolution of N$_{\rm side}$=8192. 
 These simulated skies include gravitationally lensed CMB anisotropy based on the best-fit \LCDM{} model from K11,
 a Poisson distribution of radio galaxies, 
 and Gaussian realizations of the thermal and kinetic Sunyaev-Zel'dovich (SZ) effects and cosmic infrared background (CIB).
 The lensed realizations of the CMB spectrum are generated out to $\ell=8000$ using LensPix \citep{lewis05}.  
  The Poisson radio galaxy contribution is based on the \citet{dezotti10} model for sources below the $5\,\sigma$ detection threshold in the SPT-SZ survey, and the observed counts \citep{vieira10} above that flux. 
 The shape of the thermal SZ spectrum is taken from \citet{shaw10} with an amplitude taken from \citet{reichardt12b}. 
 The kinetic SZ spectrum is based on the fiducial model in \citet{zahn12}.  
 The CIB spectrum is taken from the best-fit values in \citet{reichardt12b}.  

Unlike the simulations in K11, these simulations cover the full sky.
 The full-sky simulations make it simple to account for overlap between fields when calculating the sample variance term of the bandpower covariance matrix (see \S \ref{sec:cov}).  
 These simulations also account for any effects due to projecting from the curved sky to flat sky maps to first order in the transfer function, though these effects should be negligible, as argued in K11.

These simulated skies are observed using the SPT pointing information and then filtered and processed into maps using the same pipeline as for the real data.  
For each field, we calculate the transfer function by comparing the average power spectrum of these simulated maps to the known input spectrum using an iterative scheme \citep{hivon02}.

The transfer function is equal to $\sim 0.25$ at $\ell=650$ and reaches a plateau for $\ell \gtrsim 1200$.  
The transfer function does not reach unity at any scale due to the strong filtering of $\ell_x \lesssim 300$ modes.

\subsection{Bandpower Covariance Matrix}
\label{sec:cov}

The bandpower covariance matrix quantifies the bin-to-bin covariance of the unbiased spectrum.  
 The covariance matrix contains signal and noise terms as well as terms accounting for beam and calibration uncertainties.
 The signal term, often referred to as ``sample variance'',  is calculated from the 100 simulations described in section \ref{sec:transfer}.  
 For each simulated Healpix sky, we calculate the combined power spectrum from all fields, then measure the variance of these 100 estimates.  
 This process naturally accounts for any overlap between fields. 
 The noise term, or ``noise variance'', is estimated directly from the data using the distribution of individual cross-spectra $D^{AB}_b$ as described by \citet{lueker10}.
 The sample variance is dominant at multipoles below $\ell \lesssim 2900$.
 At smaller angular scales, the noise variance dominates.

The initial estimate of the bandpower covariance matrix  has low signal-to-noise on the off-diagonal elements. 
 As in K11, we condition the covariance matrix to reduce the impact of this uncertainty.

We must also account for the bin-to-bin covariance due to the uncertainties in the beam function $B_{\ell}$.  
We construct a ``beam correlation matrix'' for each source of beam uncertainty: 

\begin{equation}
\pmb{\rho}^{beam}_{ij} = \left(\frac{\delta D_i}{D_i}\right) \left(\frac{\delta D_j}{D_j}\right)
\end{equation}
where
\begin{equation}
\frac{\delta D_i}{D_i} = 1-\left(1+\frac{\delta B_i}{B_i}\right)^{-2}.
\end{equation}

We sum these matrices to find the full beam correlation matrix, and convert to a covariance matrix according to
\begin{equation}
\textbf{C}^{beam}_{ij} = \pmb{\rho}^{beam}_{ij}D_{i}D_{j}.
\end{equation}

\subsection{Combining Fields}
\label{sec:combine}

The analysis described in the previous sections produces 19 sets of bandpowers and covariance matrices, one from each field.  
 In either the limit of equal noise or the limit of sample variance domination, the optimal weight for each field would be its effective area (i.e., the integral of its window). 
 Since we are approximately in these limits, we use area-based weights. 
 Thus the combined bandpowers and covariance matrix are given by
\begin{equation}
D_b = \sum_{i}D_{b}^{i}w^{i}
\end{equation}
\begin{equation}
\label{eqn:combCov}
\textbf{C}_{bb^\prime} = \sum_{i}\textbf{C}_{bb^\prime}^{i}(w^{i})^2
\end{equation}
where 
\begin{equation}
w^{i} = \frac{A^i}{\sum_{i}A^i}
\end{equation}
is the area-based weight of the $i^{th}$ field.  
The area $A^i$ is the sum of the window for the $i^{th}$ field.  

We calculate the final covariance matrix as the sum of the signal plus noise covariance matrix, the beam covariance matrix, and the calibration covariance matrix. 
 For the signal and noise terms,  we combine the signal plus noise covariance matrices from all fields using Equation \ref{eqn:combCov}.
 We condition this combined covariance matrix using Equation 11 from K11.  
 For the beam covariance matrix, we take the beam covariance matrices for each year (see \S \ref{sec:cov}), and combine them into one composite beam covariance matrix using the area-based weight scheme.  
 In this step, we take care to account for the beam errors that are correlated between years.  
 Finally, we add the calibration covariance matrix, defined as  $\textbf{C}^{cal}_{ij} = \epsilon^{2}D_{i}D_{j}$, where $\epsilon=0.026$ is the $2.6\%$ uncertainty in the SPT power calibration discussed in \S \ref{sec:calibration}.

\subsection{Bandpower Window Functions}
\label{sec:windowfunc}

Bandpower window functions are necessary to compare the measured bandpowers to a theoretical power spectrum. 
The window function ${\mathcal W}^b_\ell / \ell$ is defined as

\begin{equation}
C_b^{\rm th} = ({\mathcal W}^b_\ell / \ell) C_\ell^{\rm th}.
\end{equation}

Following the formalism described in Section~\ref{sec:UnbiasedSpectra}, we can write this as

\begin{equation}
C_b^{\rm th} = (K^{-1})_{bb'} P_{b' \ell'} M_{\ell' \ell} F_\ell B_\ell^2 C_\ell^{\rm th},
\end{equation}
which implies that\footnote{Note: due to conventions in the CosmoMC package, the window functions from the publicly downloadable ``Newdat'' files should be used as follows:
\[C_b^{\rm th} = \left({\mathcal W}^b_\ell\, \frac{(\ell+0.5)}{(\ell+1)}\right)\, C_\ell^{\rm th}\]}
\begin{equation}
  {\mathcal W}^b_\ell / \ell = (K^{-1})_{bb'} P_{b' \ell'} M_{\ell' \ell} F_\ell B_\ell^2.
\end{equation}

We calculate the bandpower window functions to be used for the final spectrum measurement as the weighted average of the bandpower window functions from each field.

\section{Tests for Systematic Errors}
\label{sec:systematicTests}

It is important to verify that the data are unbiased by systematic errors.
We perform two types of tests: null tests and pipeline tests.
\subsection{Null Tests}
\label{sec:nulltests}

 As is common in CMB analyses, we check for possible systematic errors by performing a suite of null tests which are frequently referred to as jackknife tests.
 In each null test, all observations are divided into two equally sized sets based on a possible source of systematic error.  
 Difference maps are then calculated by subtracting the two sets, thus removing the astrophysical signal.  
 The power spectrum of the difference maps is calculated as described in the last section.
 This spectrum is compared to an ``expectation spectrum,'' the power we expect to see in the absence of contamination from systematic errors.  
 The expectation spectrum will generically be non-zero due to small differences in observation weights, filtering, etc, and is calculated by applying the null test to simulated maps. 
 The expected power is small ($D_{\ell} < 2 \microKsq$ at all multipoles) for all tests.

We perform six null tests:

\begin{itemize}

\item Time: Observations are ordered by time, then divided into first-half and second-half sets.  This tests for long-term temporally varying systematic effects. 

\item Scan Direction: Observations are divided into maps made from left-going scans and right-going scans.  This tests for scan-synchronous and scan-direction-dependent systematic errors.

\item Azimuthal Range: 
We split the data into observations taken at azimuths that we expect to be more or less susceptible to ground pickup.  
 These azimuth ranges are determined from maps of the 2009 data made using ``ground-centered'' (Azimuth/Elevation) coordinates in which ground pickup adds coherently, as opposed to the usual ``sky-centered'' (R.A./decl.) coordinates.
 We use the ground-centered maps that were made for the analysis presented by K11.
 Although we detect emission from the ground on large scales ($\ell\sim50$) in these ground-centered maps, this is not expected to bias our measurement; 
 the amplitude of the ground pickup is significantly lower on the smaller angular scales for which the bandpowers are being reported, and the observations for a given field are distributed randomly in azimuth.
 We use the azimuth-based null test to verify this assertion.

\item Moon:  Observations are divided into groups based on when the Moon was above and below the horizon.

\item Sun:  Observations are divided into groups based on when the Sun was above and below the horizon.  In this test, we only include fields in which more than $25\%$ of the observations were taken with the Sun above the horizon.

\item Summed Bolometer Weights:  We calculate the sum of all bolometer weights during each observation and order maps based on this sum.  This tests for bias introduced by incorrectly weighting observations or incomplete coverage in some maps.

\end{itemize}

For each test, the $\chi^2$ of the residual power is calculated relative to the expectation spectrum in five bins with $\delta \ell=500$.  
 We calculate the probability to exceed (PTE) this value of $\chi^2$ for five degrees of freedom.  
 All null tests had reasonable PTE's, as listed below, with the exception of the Azimuthal Range null test, which produced a low PTE for the original set of observations.
 This was interpreted as evidence for some ground contamination.
 This interpretation was tested by cutting several sets of 5\% of the data, and re-calculating the Azimuthal Range null test.  
 Removing random 5\% sets of the data did not change the failure of the Azimuthal Range null test. 
 However, cutting the 5\% of the observations from each field with the highest expected ground contamination resulted in passing the Azimuthal Range null test.
 Thus this cut was included with the other observation cuts, as described in \S~\ref{sec:fields_and_obs}.
 It is worth noting that the Azimuthal Range null test is the worst-case scenario for ground pickup; this null test systematically aligns azimuth ranges to maximize the ground contamination. 
 In the analysis of the power spectrum of the sky signal, the ground signal will add incoherently as the azimuth changes, thus reducing the power from ground contamination to a much lower level than in this null test.
 
We find a flat distribution of PTEs ranging from 0.02 to 0.99 for individual fields.
 The combined PTEs for the Time, Scan Direction, Azimuthal Range, Moon, Sun, and Summed Bolometer Weights are 0.26, 0.14, 0.17, 0.13, 0.30, 0.63, respectively. 
 It is important to note that these null tests are extremely conservative for the SPT power spectrum where the uncertainties are sample variance dominated over most of the range reported in this work.
 It is possible to have a failure in these null tests without a significant impact on the final power spectrum.
 The measured power $D_{\ell}$  in each null test was less than $2.2\, \mu K^2$ in all bins.

Though we cannot perform a direct year-to-year null test because the data for any given field was taken within a single year,
 we have verified that the spectra from different years (and therefore different fields, including those used in K11) are consistent within the uncertainties of cosmic variance.

\subsection{Pipeline Tests}
\label{sec:pipelineTests}
We test the robustness of our pipeline with simulations.
 In these tests, we create simulated maps with an input spectrum that differs from the \LCDM{} model spectrum assumed in the calculation of the transfer function.
 We then use our full pipeline to calculate the power spectrum of these simulated maps, and compare this spectrum with the input spectrum.
 We looked at three categories of modifications:
\begin{itemize}
  \item A slope was added to the best-fit \LCDM{} spectrum from K11.  This tests how well we can measure the slope of the damping tail.
  \item An additional Poisson point-source power term was added (see \S~\ref{sec:model}).
  \item The input spectrum was shifted by $\delta_{\ell}=10$.  This tests how well we can measure the locations of the acoustic peaks, and therefore $\theta_s$.
\end{itemize}
In all cases, we recover the input spectrum to well within our uncertainties.

 We thus find no significant evidence for systematic contamination of SPT bandpowers.
\section{Bandpowers}
\label{sec:bandpowers}
\begin{figure*}
\begin{center}
    \includegraphics[width=0.98\textwidth]{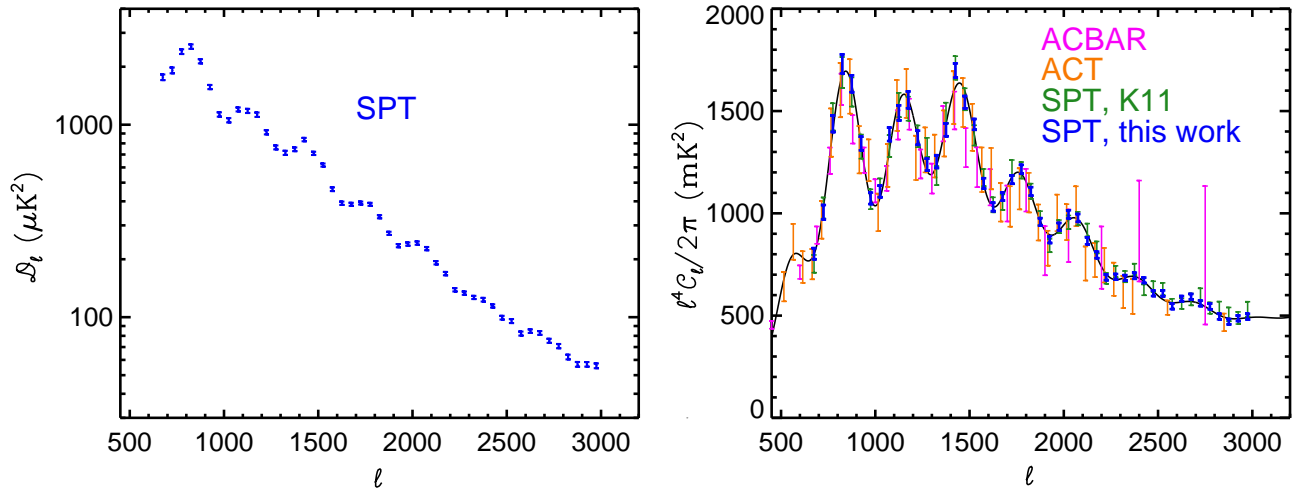}
\end{center}
\caption{\textbf{\textit{Left panel:}} The SPT power spectrum. 
  The leftmost peak at $\ell \sim800$ is the third acoustic peak. 
  \textbf{\textit{Right panel:}} 
A comparison of the new SPT bandpowers with other recent measurements of the CMB damping tail from ACBAR \citep{reichardt09a}, ACT \citep{das11}, and SPT (K11). 
  Note that the point source masking threshold differs between these experiments which can affect the power at the highest multipoles. 
  In order to highlight the acoustic peak structure of the damping tail, we plot the bandpowers in the right panel as $\ell^4\clnospace/(2\pi)$, as opposed to $D_{\ell}=\ell(\ell+1)\clnospace/(2\pi)$ in the left panel. 
  The solid line shows the theory spectrum for the \LCDM model + foregrounds that provides the best fit to the SPT+\wseven{} data. 
  The bandpower errors shown in these plots contain sample and noise variance terms only; they do not include beam or calibration uncertainties.}
\label{fig:dl_all}
\end{figure*}

\begin{figure*}
\begin{center}
    \includegraphics[width=0.98\textwidth]{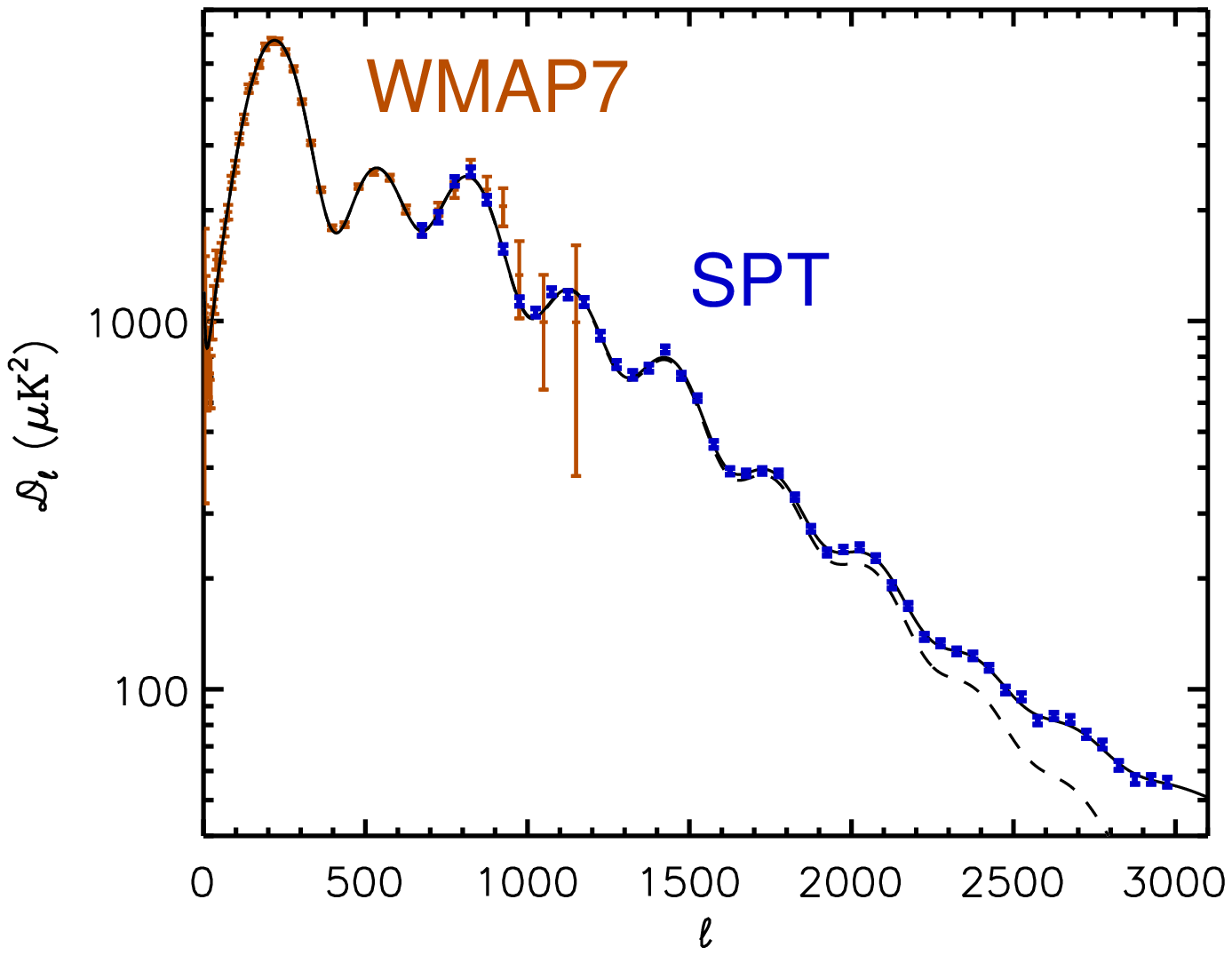}
\end{center}
\caption{
The SPT bandpowers (\textbf{blue}), \wseven{} bandpowers (\textbf{orange}), and the lensed \LCDMnospace+foregrounds theory spectrum that provides the best fit to the SPT+\wseven{} data shown for the CMB-only component (\textbf{dashed line}), and the CMB+foregrounds spectrum (\textbf{solid line}).
 As in Figure~\ref{fig:dl_all}, the bandpower errors shown in this plot  do not include beam or calibration uncertainties.}
\label{fig:ps}
\end{figure*}

Following the analysis presented in \S \ref{sec:ps}, we measure the CMB temperature anisotropy power spectrum from 2540 deg$^2$ of sky observed by the SPT between 2008 and 2011.  
 We report bandpowers in bins of $\delta \ell = 50$ between $650 < \ell < 3000$.
 The bandpowers and associated errors are listed in Table~\ref{tab:dls} and shown in Figures~\ref{fig:dl_all} and \ref{fig:ps}.  
 The bandpowers, covariance matrix, and window functions are available for download on the SPT website.\footnote{http://pole.uchicago.edu/public/data/story12/}

These bandpowers clearly show the third to ninth acoustic peaks.
 As Figure~\ref{fig:ps} demonstrates, the anisotropy power measured by this analysis (at 150 GHz, with a 50mJy point-source cut) is dominated by primary CMB, with secondary anisotropy and foregrounds contributing significantly only at the highest multipoles.
 These bandpowers provide the most precise measurement to date of the CMB power spectrum over the entire multipole range presented in this analysis.

\begin{table*}[ht!]
\begin{center}
\caption[SPT Bandpowers and Bandpower Errors]{SPT Bandpowers and Bandpower Errors}
\small
\begin{tabular}{cr|rr @{\hskip 10 mm}| @{\hskip 10 mm}cr|rr}
\hline\hline
\rule[-2mm]{0mm}{6mm}
$\ell$ range&$\ell_{\rm eff}$ &$D_{\ell}$ [$\microKsq$]& $\sigma$ [$\microKsq$] & $\ell$ range&$\ell_{\rm eff}$ &$D_{\ell}$ [$\microKsq$]& $\sigma$ [$\microKsq$]\\
\hline
 651 -   700 &    671 & 1786.2 & 59.5 &   1851 -  1900 &   1865 &  276.1 &  5.2 \\
 701 -   750 &    720 & 1939.3 & 66.9 &   1901 -  1950 &   1915 &  238.2 &  4.5 \\
 751 -   800 &    770 & 2426.4 & 67.2 &   1951 -  2000 &   1966 &  242.8 &  4.6 \\
 801 -   850 &    820 & 2577.1 & 68.3 &   2001 -  2050 &   2015 &  245.8 &  4.9 \\
 851 -   900 &    870 & 2162.3 & 53.8 &   2051 -  2100 &   2064 &  229.9 &  4.5 \\
 901 -   950 &    920 & 1588.8 & 39.0 &   2101 -  2150 &   2114 &  194.2 &  3.8 \\
 951 -  1000 &    969 & 1144.3 & 29.6 &   2151 -  2200 &   2164 &  170.6 &  3.4 \\
1001 -  1050 &   1019 & 1068.0 & 27.2 &   2201 -  2250 &   2213 &  140.5 &  2.8 \\
1051 -  1100 &   1069 & 1215.7 & 28.5 &   2251 -  2300 &   2265 &  135.0 &  2.6 \\
1101 -  1150 &   1118 & 1193.8 & 29.1 &   2301 -  2350 &   2313 &  128.3 &  2.4 \\
1151 -  1200 &   1169 & 1141.1 & 29.8 &   2351 -  2400 &   2364 &  124.8 &  2.7 \\
1201 -  1250 &   1218 &  924.8 & 23.1 &   2401 -  2450 &   2413 &  115.8 &  2.2 \\
1251 -  1300 &   1269 &  771.7 & 17.9 &   2451 -  2500 &   2462 &  100.7 &  2.2 \\
1301 -  1350 &   1318 &  723.1 & 17.7 &   2501 -  2550 &   2512 &   96.7 &  2.3 \\
1351 -  1400 &   1367 &  754.6 & 16.6 &   2551 -  2600 &   2562 &   83.3 &  2.0 \\
1401 -  1450 &   1417 &  847.3 & 17.0 &   2601 -  2650 &   2613 &   85.7 &  1.8 \\
1451 -  1500 &   1468 &  718.7 & 13.8 &   2651 -  2700 &   2663 &   83.9 &  1.9 \\
1501 -  1550 &   1517 &  625.0 & 11.3 &   2701 -  2750 &   2712 &   76.4 &  1.8 \\
1551 -  1600 &   1567 &  468.1 & 10.2 &   2751 -  2800 &   2761 &   71.7 &  1.8 \\
1601 -  1650 &   1617 &  395.7 &  7.9 &   2801 -  2850 &   2811 &   62.9 &  1.7 \\
1651 -  1700 &   1666 &  390.7 &  7.0 &   2851 -  2900 &   2860 &   57.6 &  1.6 \\
1701 -  1750 &   1717 &  396.6 &  6.9 &   2901 -  2950 &   2910 &   57.6 &  1.6 \\
1751 -  1800 &   1766 &  390.7 &  6.9 &   2951 -  3000 &   2961 &   56.6 &  1.6 \\
1801 -  1850 &   1815 &  336.7 &  6.2 & & & \\
\hline
\end{tabular}
\label{tab:dls}
\tablecomments{
The $\ell$-band range, weighted multipole value $\ell_{\rm eff}$, bandpower $D_{\ell}$, and associated bandpower uncertainty $\sigma$ of the SPT power spectrum.  
The errors are the square-root of the diagonal elements of the covariance matrix, and do not include beam or calibration uncertainties.
}
\end{center}
\end{table*}

\section{Cosmological Constraints}
\label{sec:constraints}

The SPT bandpowers are high signal-to-noise measurements of the CMB temperature anisotropy over a large range of angular scales, and can be used to perform sensitive tests of cosmological models.
 In this section, we present the constraints these bandpowers place on cosmological models.
 We first constrain the standard \LCDM{} cosmological model.
 Next, we extend this model to constrain the amplitude of gravitational lensing of the CMB.
 We then consider models with free spatial curvature and constrain the mean curvature of the observable universe.
 We finally consider tensor perturbations and discuss the implications of our observations for simple models of inflation. 
 A wider range of cosmological models are tested in a companion paper H12.

We parametrize the \LCDM{} model with six parameters: the baryon density $\Omega_b h^2$, the density of cold dark matter $\Omega_c h^2$, the optical depth to reionization $\tau$, the angular scale of the sound horizon at last scattering $\theta_s$, the amplitude of the primordial scalar fluctuations (at pivot scale $k_0=0.05$ Mpc$^{-1}$) $\Delta^2_{R}$, and the spectral index of the scalar fluctuations \ns{}.  
 With the exception of \S~\ref{sec:curvature}, we consider only flat-universe models where the mean curvature of the observable universe $\omk=0$.
 In addition to the six parameters described above, we report several derived parameters that are calculated from the six \LCDM{} parameters.
 These are the dark energy density $\Omega_\Lambda$, the Hubble constant \ho{} in units of km s$^{-1}$ Mpc$^{-1}$, the current amplitude of linear matter fluctuations $\sigma_8$ on scales of $8\,h^{-1}$ Mpc, the redshift of matter-radiation equality $z_{\rm EQ}$, and a hybrid distance ratio reported by baryon acoustic oscillation (BAO) experiments at two different redshifts $r_s/D_v(z=0.35)$ and $r_s/D_v(z=0.57)$, where $r_s$ is the comoving sound horizon size at the baryon drag epoch, $D_V(z) \equiv [(1 + z)^2 D^2_A(z)cz/H(z)]^{1/3}$, $D_A(z)$ is the angular diameter distance, and $H(z)$ is the Hubble parameter.

\subsection{Foreground treatment}
\label{sec:model}

We marginalize over three foreground terms in all parameter fitting. 
The total foreground power, $D_\ell^{\rm fg}$, can be expressed as:
\begin{equation}
D_\ell^{\rm fg} = D_\ell^{\rm gal} + D^{\rm SZ}_\ell
\end{equation}
These two terms represent the following:

\begin{itemize}

\item \textit{Power from galaxies ($D_\ell^{\rm gal}$)},  which can be subdivided into a clustering term and a Poisson term. 
The Gaussian priors used by K11 are applied to the amplitude of each term at $\ell=3000$. 
For the clustering term, the prior is $D^{\rm CL}_{3000} = 5.0 \pm 2.5\, \microKsq$, based on measurements by  \cite{shirokoff11}. 
The angular dependence of the clustering term is $D_\ell^{\rm CL} \propto \ell^{0.8}$, which has been modified from that assumed by K11 to agree better with recent measurements (e.g., \citet{addison12, reichardt13}).  
For the Poisson term, the prior is  $D^{\rm PS}_{3000} = 19.3 \pm 3.5\, \microKsq$. 
This is based on the power from sources with $S_{\rm 150 GHz} < 6.4~\rm{mJy}$, as measured in \cite{shirokoff11}, and the power from sources with $6.4~\rm{mJy}$ $< S_{\rm 150 GHz} < 50~\rm{mJy}$, as measured in \cite{vieira10} and \cite{marriage11a}. 
The Poisson term is constant in $C_{\ell}$ and thus varies as $D^{\rm PS}_\ell \propto \ell^2$.

\item \textit{SZ power ($D^{\rm SZ}_\ell$).} 
The thermal and kinetic SZ effects are expected to contribute to the observed CMB temperature anisotropy. 
Both effects are expected to have similar power spectrum shapes over the angular scales relevant to this analysis.
Therefore we adopt a single template to describe both effects. 
The chosen template is the thermal SZ model from \citet{shaw10}. 
We set a Gaussian prior on the amplitude of this term of $D^{\rm SZ}_{3000} = 5.5 \pm 3.0\, \microKsq$, as measured in \cite{shirokoff11}.
This amplitude is defined at 153 GHz, corresponding to the effective SPT band center.

\end{itemize}

We have tested that all parameter constraints are insensitive to the details of the assumed foreground priors; 
we have completely removed the priors on the amplitudes of the foreground terms and re-calculated the best-fit \LCDM\ model, and find all \LCDM{} parameters shift by less than $0.06\,\sigma$.
Additionally, we see no evidence for significant correlations between the foreground and cosmological parameters.

In the above model, we have not accounted for the emission from cirrus-like dust clouds in the Milky Way.  
 Repeating the calculation performed in K11, we cross-correlate the SPT maps with predictions for the galactic dust emission at $150$~GHz in the SPT fields using model 8 of \citet{finkbeiner99}.  We use this cross-correlation to estimate the power from the galactic dust in the SPT fields, and find that it is small compared to the primary CMB power and the SPT bandpower errors.
 Specifically, subtracting this cirrus power can be balanced by moving the foreground terms by amounts that are small compared to their priors, so that the change in $\chi^2$ is less than $0.1\,\sigma$.
 Thus we conclude that galactic dust does not significantly contaminate the SPT power spectrum.

\subsection{Estimating Cosmological Parameters}
Our baseline model contains nine parameters: six for the primary CMB and three for foregrounds.
 We explore the nine-dimensional parameter space using a Markov Chain Monte Carlo (MCMC) technique \citep{christensen01} implemented in the CosmoMC\footnote{http://cosmologist.info/cosmomc/} \citep{lewis02b} software package.
 For reasons of speed, we use PICO\footnote{\url{https://sites.google.com/a/ucdavis.edu/pico}} \citep{fendt07a,fendt07b}, trained with CAMB\footnote{http://camb.info/ (January 2012 version)} \citep{lewis99}, to calculate the CMB power spectrum. 
 We have trained PICO for a ten-parameter model that includes \LCDM\ as well as several extensions.
 We use PICO when working with any subset of this model space, and CAMB for all other extensions to \LCDM. 
 The effects of gravitational lensing on the power spectrum of the CMB are calculated using a cosmology-dependent lensing potential \citep{lewis06}. 
 To sample the posterior probability distribution in regions of very low probability, we run ``high-temperature'' chains in which the true posterior, $P$, is replaced in the Metropolis Hastings algorithm by $P_T = P^{1/6}$.  
 This allows the chain to sample the parameter space more broadly.  
 We recover the correct posterior from the chain by importance sampling each sample with weight $P/P_T$.

\subsection{Goodness of fit to the \LCDM{} Model}
We quantify the goodness of fit of the \LCDM{} model to the SPT bandpowers by finding the spectrum which best fits the SPT bandpowers and calculating the reduced $\chi^2$ for the SPT data.
 The reduced $\chi^2$ for the SPT data is 45.9/39 (PTE=0.21), thus the \LCDM{} model is a good fit to the SPT bandpowers.
 In H12, we consider several extensions to the \LCDM{} model, and find that the data show some preference for several of those extensions.

\subsection {External Datasets}
\label{subsec:datasets}

In this work, we focus on parameter constraints from the CMB data, sometimes in conjunction with measurements of the Hubble constant ($H_0$) or the BAO feature. 
 For CMB measurements, we use the SPT bandpowers presented here as well as the \wmap bandpowers presented in \wseven{}.
 For \ho{} measurements, we use the low-redshift measurement from \citet{riess11}.
 For the BAO feature, we use a combination of three measurements at different redshifts:  the WiggleZ survey covering the redshift range $0.3 < z < 0.9$ \citep{blake11}, the SDSS-II survey (DR7)  covering $0.16<z<0.44$ \citep{padmanabhan12}, and the BOSS survey covering $0.43<z<0.7$ \citep{anderson12}.

Before combining the CMB, \ho{}, and BAO datasets, we check their relative consistency within the \LCDM model.  
We quantify this consistency by calculating the $\chi^2_{\rm min}$ using a reference dataset (e.g., CMB) and comparing it to the $\chi^2_{\rm min}$ obtained using a new dataset (e.g., CMB+\ho{}).  For example, $\chi^2_{\rm min, [CMB+\ho{}]} - \chi^2_{\rm min, [CMB]}$=0.08.  
The probability to exceed this $\Delta\chi^2$ given the one new degree of freedom provided by the \ho{} measurement is 0.78, corresponding to an effective Gaussian significance of 0.3$\,\sigma$.  Using this metric, we find that
\begin{itemize}
\item{CMB and \ho{} differ by 0.3$\,\sigma$.}
\item{CMB and BAO differ by 1.5$\,\sigma$.}
\item{(CMB+BAO) and \ho{} differ by 1.8$\,\sigma$.}
\item{(CMB+\ho{}) and BAO differ by 2.1$\,\sigma$.}
\end{itemize}

There is some tension between these datasets in the context of the \LCDM model. 
 This could be evidence for a departure from \LCDM, a systematic error in one or more of the data sets, or simply a statistical fluctuation. 
 We assume the uncertainties reported for each of the datasets are correct and combine them to produce many of the results presented here. 

\subsection {SPT-only \LCDM{} constraints}
\label{subsec:w7_s12tau}

We begin by examining parameter constraints from the SPT bandpowers alone. 
 The SPT-only parameter constraints provide an independent test of \LCDM{} cosmology and allow for consistency checks between the SPT data and other datasets. 
 Because the scalar amplitude $\Delta^2_{R}$ and the optical depth $\tau$ are completely degenerate for the SPT bandpowers, we impose a \wseven-based prior of $\tau = 0.088 \pm 0.015$ for the SPT-only constraints. 

\begin{figure*}
\begin{center}
    \includegraphics[angle=90,width=0.98\textwidth]{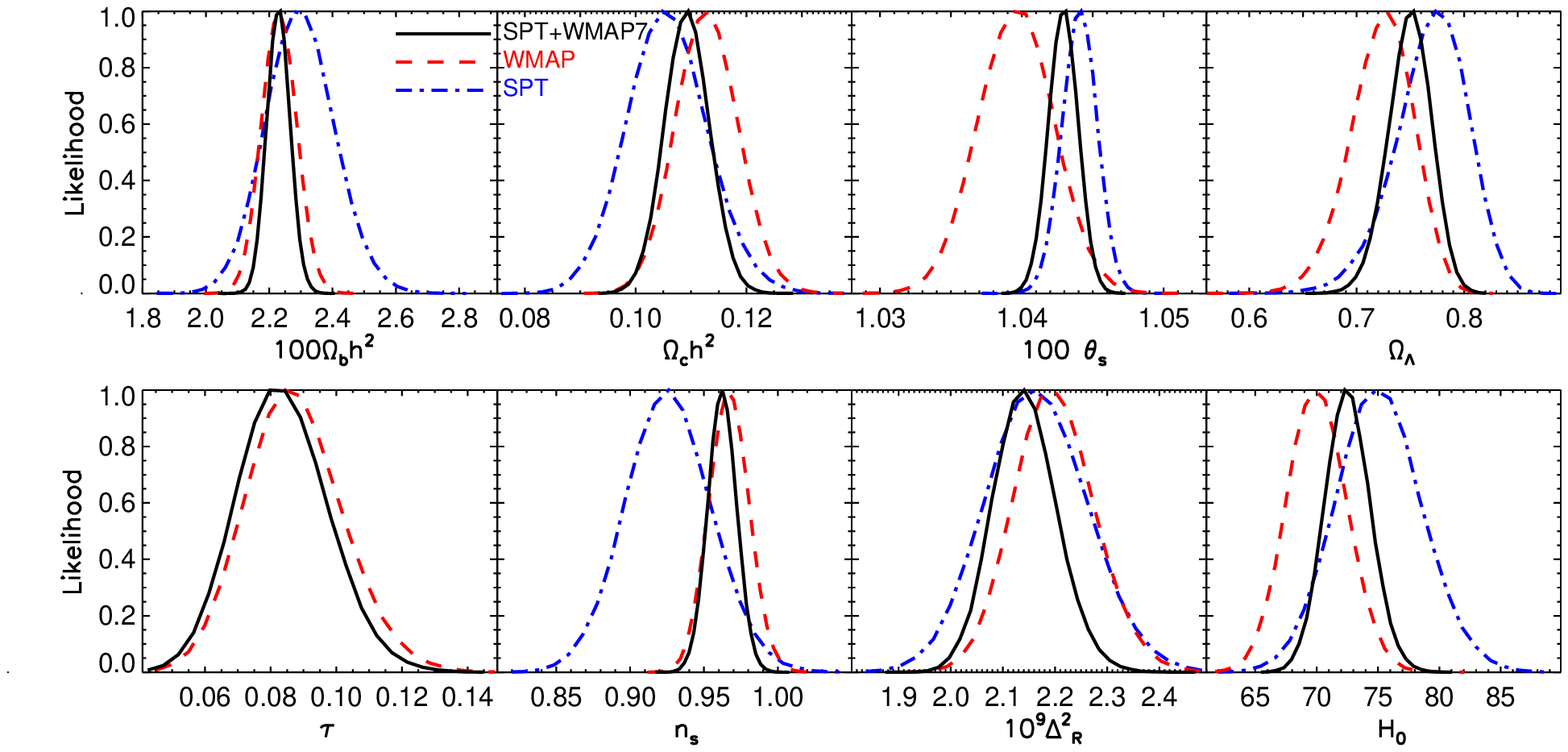}
\end{center}
\caption{
The one-dimensional marginalized likelihoods of the six parameters of the \LCDM{} model, plus two derived parameters: the dark energy density $\Omega_{\Lambda}$ and the Hubble constant $H_0$.
 The constraints are shown for the SPT-only (\textbf{blue dot-dashed lines}), \wseven-only (\textbf{red dashed lines}), and SPT+\wseven{}  (\textbf{black solid lines}) datasets.
With the exception of $\tau$, the SPT bandpowers constrain the parameters approximately as well as \wseven{} alone. 
In particular, the SPT bandpowers measure the angular sound horizon $\theta_s$ extremely well because they measure seven acoustic peaks.
  In the SPT-only constraints, the \wseven{} measurement of $\tau$ has been applied as a prior; because of this we do not plot an SPT-only line on the $\tau$ plot.}
\label{fig:like1d_s12tau_w7}
\end{figure*}

We present the constraints on the \LCDM{} model from SPT and \wseven{} data in columns two to four of Table~\ref{tab:lcdm}.
 As shown in Figure~\ref{fig:like1d_s12tau_w7}, the SPT bandpowers (including a prior on $\tau$ from \wseven) constrain the \LCDM{} parameters approximately as well as \wseven.
 The SPT and \wseven{} parameter constraints are consistent for all parameters; $\theta_s$ changes the most significantly among the five free \LCDM parameters, moving by $1.5\, \sigma$ and tightening by a factor of 2.2 from \wseven{} to SPT.
 The SPT bandpowers measure $\theta_s$ extremely well by virtue of the sheer number of acoustic peaks -- seven -- measured by the SPT bandpowers. 
 The SPT constraint on \ns{} is broader than the constraint from \wseven{} due to the fact that \wseven{} probes a much greater dynamic range of angular scales.
 Degeneracies with \ns{} degrade the SPT constraints on $\Delta_R^2$, the baryon density and, to a lesser extent, the dark matter density.

\subsection{Combined \LCDM{} constraints}
\label{subsec:w7+s12}

\begin{table*}
\begin{small}
\begin{center}
\begin{threeparttable}
\caption{\LCDM{} Parameter Constraints from the CMB and external datasets}
\footnotesize
\begin{tabular}{c | c | c | c | c | c | c}
\hline \hline
Parameter & \wseven{} & SPT\tablenotemark{(a)} & CMB &  CMB+$H_0$ &  CMB+BAO & CMB+$H_0$+BAO \\
 & & & \footnotesize{(SPT+\wseven)} &  & & \\
\hline
\multicolumn{6}{l}{Baseline parameters} \\
$100\,\Omega_bh^2$       & $2.231 \pm 0.055$ &     $2.30  \pm 0.11 $ &     $2.229 \pm 0.037$ &     $2.233 \pm 0.035$ &     $2.204 \pm 0.034$ &     $2.214 \pm 0.034$ \\     
$\Omega_ch^2$           & $0.1128 \pm 0.0056$ &   $0.1056 \pm 0.0072$ &   $0.1093 \pm 0.0040$ &   $0.1083 \pm 0.0033$ &   $0.1169 \pm 0.0020$ &   $0.1159 \pm 0.0019$ \\   
$10^{9} \deltaR$        & $2.197 \pm 0.077$ &     $2.164 \pm 0.097$ &     $2.142 \pm 0.061$ &     $2.138 \pm 0.062$ &     $2.161 \pm 0.057$ &     $2.160 \pm 0.057$ \\     
$n_s$                   & $0.967  \pm 0.014 $ &   $0.926  \pm 0.029 $ &   $0.9623 \pm 0.0097$ &   $0.9638 \pm 0.0090$ &   $0.9515 \pm 0.0082$ &   $0.9538 \pm 0.0081$ \\   
$100\,\theta_s$          & $1.0396  \pm 0.0027 $ & $1.0441  \pm 0.0012 $ & $1.0429  \pm 0.0010 $ & $1.0430  \pm 0.0010 $ & $1.04215 \pm 0.00098$ & $1.04236 \pm 0.00097$ \\ 
$\tau$                  & $0.087 \pm 0.015$ &     $0.087 \pm 0.015$ &     $0.083 \pm 0.014$ &     $0.084 \pm 0.014$ &     $0.076 \pm 0.012$ &     $0.077 \pm 0.013$ \\     
\hline
\multicolumn{7}{l}{        Derived parameters\tablenotemark{(b)}} \\
$\Omega_\Lambda$        & $0.724 \pm 0.029 $ &   $0.772  \pm 0.033 $ &   $0.750  \pm 0.020 $ &   $0.755  \pm 0.016 $ &   $0.709  \pm 0.011 $ &   $0.7152 \pm 0.0098$ \\   
\ho                     & $70.0 \pm  2.4 $ &     $75.0  \pm  3.5 $ &     $72.5  \pm  1.9 $ &     $73.0 \pm  1.5 $ &     $69.11 \pm  0.85$ &     $69.62 \pm  0.79$ \\      
$\sigma_8$              & $0.819 \pm 0.031$ &     $0.772 \pm 0.035$ &     $0.795 \pm 0.022$ &     $0.791 \pm 0.019$ &     $0.827 \pm 0.015$ &     $0.823 \pm 0.015$ \\     
$z_{\rm EQ}$            & $3230 \pm  130 $ &     $3080  \pm  170 $ &     $3146  \pm   95 $ &     $3124  \pm   78 $ &     $3323  \pm   50 $ &     $3301  \pm   47 $ \\     
$100\frac{r_s}{D_V}(z=0.35)$       & $11.43 \pm  0.37$ &     $12.15 \pm  0.55$ &     $11.81 \pm  0.29$ &     $11.89 \pm  0.24$ &     $11.28 \pm  0.12$ &     $11.35 \pm  0.12$ \\     
$100\frac{r_s}{D_V}(z=0.57)$       & $ 7.58 \pm  0.21 $ &   $ 7.98  \pm  0.31 $ &   $ 7.80  \pm  0.16 $ &   $ 7.84  \pm  0.13 $ &   $ 7.505 \pm  0.068$ &   $ 7.545 \pm  0.065$ \\ 
\hline
\end{tabular}
\label{tab:lcdm}
\end{threeparttable}
\begin{tablenotes}[para]
\textbf{Notes:} The constraints on cosmological parameters from the \LCDM model, given five different combinations of datasets.  
  We report the median of the likelihood distribution and the symmetric 68.3\% confidence interval about the mean. \\ \\
$^{\rm (a)}$\, We impose a \wseven-based prior of $\tau = 0.088 \pm 0.015$ for the SPT-only constraints. \\
$^{\rm (b)}$\, Derived parameters are calculated from the baseline parameters in CosmoMC. They are defined at the end of \S~\ref{sec:model}.\\
\end{tablenotes}
\end{center}
\end{small}
\end{table*}

Next, we present the constraints on the \LCDM{} model from the combination of SPT and \wseven{} data.
 As previously mentioned, we will refer to the joint SPT+\wseven{} likelihood as the CMB likelihood. 
 We then extend the discussion to include constraints from CMB data in combination with BAO and/or \ho{} data.

We present the CMB constraints on the six \LCDM{} parameters in the fourth column of Table~\ref{tab:lcdm}.  
 Adding SPT bandpowers to the \wseven{} data tightens these parameter constraints considerably relative to \wseven{} alone.
 Of these parameters, the constraint on $\theta_s$ sees the largest improvement; adding SPT data decreases the uncertainty on $\theta_s$ by a factor of 2.70 relative to \wseven{} alone.
 Constraints on on $\Omega_bh^2$, $\Omega_ch^2$, and $\Omega_\Lambda$ tighten by factors of 1.49, 1.40, and 1.45, respectively.
 For comparison, the addition of the K11 bandpowers to \wseven{} led to improvements of 1.33, 1.16, and 1.11, respectively.
 Finally, the constraint on the scalar spectral index tightens by a factor of 1.44 to give $n_s<1.0$ at $3.9\,\sigma$.

The preferred values for $\Omega_b h^2$ and $\Omega_c h^2$ for \wseven{} do not shift significantly with the addition of the SPT data and, therefore, neither does the sound horizon, $r_s$, which depends only on these parameters in the \LCDM model.
 Thus the shift in $\theta_s= r_s/D_A$, driven by the SPT acoustic peak locations, must lead to a shift in $D_A$.  
 Shifting $D_A$ requires shifting $\Omega_\Lambda$ (or, equivalently, \ho).  
 This shift in $\Omega_\Lambda$ and \ho{} can be seen in Figure~\ref{fig:like1d_s12tau_w7}.

We explore how the SPT+\wseven{} constraints on the \LCDM{} model change if different $\ell$-ranges of the SPT data are used in Appendix~\ref{sec:lrange}.

We show the parameter constraints after adding the \ho{} and/or BAO data to the CMB data in the last three columns of Table~\ref{tab:lcdm}.
 Combining the CMB bandpowers with this low-redshift information tightens the constraints on $\Omega_ch^2$ and $\Omega_\Lambda$ by a further factor of 1.2 or 2 for CMB+\ho{} and CMB+BAO, respectively, with smaller but significant improvements to other parameters.
 Of special note is the constraint on the scalar spectral index which tightens to $\ns=\nsCmbHo$ for the CMB+\ho{} dataset, $\nsCmbBao$ for the CMB+BAO dataset, and $\nsCmbHoBao$ for the CMB+$H_0$+BAO dataset.
 These constraints correspond to a preference for $n_s < 1$ at $4.0\,\sigma$, $6.1\,\sigma$, and $5.9\,\sigma$ respectively, for these three data combinations.
 This is the most significant reported measurement of $n_s \ne 1$ to date. 
 See \S~\ref{sec:inflation} for a more detailed discussion of constraints on \ns{}.

\subsubsection{Consistency of \LCDM{} constraints}
Comparing the best-fit cosmological model with K11, all \LCDM parameters are consistent at $< 1\,\sigma$ with the exception of $\theta_s$, which shifts up by $1.0\,\sigma$.  
It is not surprising that the most significant shift is seen in $\theta_s$; of the six \LCDM parameters, SPT data has the strongest effect on the $\theta_s$ constraint, as can be seen in Figure~\ref{fig:like1d_s12tau_w7}.
Thus the results in this paper are consistent with those from K11.

The $\sigma_8$ constraints presented here are consistent with previous measurements.                         
The SPT-only and \wseven-only values are consistent at $\sim 1\,\sigma$, with the SPT data preferring a lower value.
The ACT+\wseven{} constraint, $0.813 \pm 0.028$ \citep{dunkley11}, is consistent with that from SPT+\wseven{}, though we note that the \wseven{} data are used in both.
Comparing to X-ray measurements of cluster abundance, we re-scale the $\sigma_8$ constraint from \citet{vikhlinin09} to the SPT+\wseven{} value of $\Omega_M$ to find $\sigma_8=0.813 \pm 0.027$, which is again consistent with our measured values.
Optical and SZ-based surveys give comparable and consistent constraints, for example, \cite{rozo10} and \citet{reichardt13}.
Finally, SPT gravitational lensing measurements are consistent; \citet{vanengelen12} found $\sigma_8 = 0.810 \pm 0.026$ (\wseven{}+SPT$_{\rm Lensing}$).
Further discussion of $\sigma_8$ constraints, particularly in the context of the \LCDM+\sumnu{} model, can be found in H12.

\subsection{Gravitational Lensing}
\label{sec:lensing}

\begin{figure}
\begin{center}
    \includegraphics[width=0.48\textwidth]{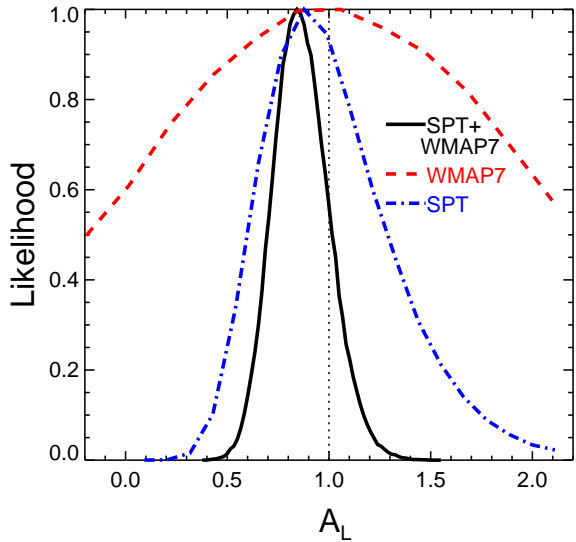}
\end{center}
\caption{The SPT bandpowers allow a significant detection of gravitational lensing through the effective smoothing of the acoustic peaks. 
Here, we show the one-dimensional likelihood function for $A_L$, a rescaling parameter for the gravitational lensing potential power spectrum ($C_\ell^{\phi\phi} \rightarrow A_{L} C_\ell^{\phi\phi}$).
The SPT+\wseven{} data lead to a $\alensCdfCmb\,\sigma$ detection of CMB lensing, the most significant detection to date. 
}
\label{fig:alens}
\end{figure}

As CMB photons travel from the surface of last scattering to the Earth, their paths are deflected by gravitational interactions with intervening matter.  
 This gravitational lensing encodes information about the distribution of matter along the line of sight, providing a probe of the distance scale and growth of structure at intermediate redshifts ($0.5 \lesssim z \lesssim 4$).
 Lensing distorts the CMB anisotropy by shifting the apparent position of CMB photons on the sky, with typical deflection angles of 2.5 arcminutes which are coherent over degree scales. 
 This process mixes power between multipoles in the CMB temperature power spectrum, which smooths the acoustic peak structure and increases the power in the damping tail at small angular scales (see \citealt{lewis06} for a review). 
  
The \LCDM{} model already includes the effects of gravitational lensing.
 To quantify the sensitivity of the SPT bandpowers to gravitational lensing, we extend the \LCDM{} model to include one additional free parameter, $A_L$ \citep{calabrese08}, which rescales the lensing potential power spectrum, $C_\ell^{\phi\phi}$, according to
\begin{equation}
C_\ell^{\phi\phi} \rightarrow A_{L} C_\ell^{\phi\phi}.
\end{equation} 
 We re-calculate $C_\ell^{\phi\phi}$ in a cosmology-dependent manner at each point in the MCMC. 
 In effect, the $A_L$ parameter modulates the amplitude of gravitational lensing.
 Setting $A_L = 1$ corresponds to the standard theoretical prediction and recovers the standard \LCDM model, while setting $A_L = 0$ corresponds to no gravitational lensing.
In the parameter fits, the range of $A_L$ is allowed to extend well above 1 and below 0. 

The first detection of gravitational lensing in the CMB used lensing-galaxy cross-correlations \citep{smith07, hirata08}, and subsequent papers using this technique have achieved higher signal-to-noise detections \citep{bleem12b, sherwin12}.

The impact of lensing on the CMB power spectrum has been detected in combinations of \wmap with ACBAR \citep{reichardt09a}, \wmap with ACT \citep{das11b}, and \wmap with SPT (K11). 
 Using this effect, \cite{das11b} found $A_L = 1.3^{+0.5}_{-0.5}$ at 68\% confidence.
K11 found that the constraint on $A^{0.65}$ had the most Gaussian shape and thus reported $A^{0.65}_{L} = 0.94 \pm 0.15$, a $\sim5\,\sigma$ detection of lensing.

CMB lensing has also been detected through the CMB temperature four-point function in ACT \citep{das11} and SPT \citep{vanengelen12} data. 
 \citet{das11} used ACT data to measure\footnote{In \citet{das11}, $A_L$ is calculated as the best-fit amplitude to the lensing potential in the cosmological model with the maximum likelihood \LCDM{} parameters. 
 This is in contrast to what was done for the CMB temperature power spectrum measurements of $A_L$, where constraints on $A_L$ have been marginalized over cosmological parameters.
 The corresponding maximum likelihood measure from \citet{vanengelen12} is $A_L^{\rm ML} = 0.86 \pm 0.16$.}
$A_L^{\rm ML} =1.16 \pm 0.29$.
 In \citet{vanengelen12}, the SPT four-point analysis was applied to a subset of the data used in this work to measure $\alens = 0.90 \pm 0.19$, which was previously the most significant detection of CMB lensing to date, ruling out no lensing at $6.3\,\sigma$.

We determine the significance of the observed CMB lensing by constraining $A_L$ with the measured CMB power spectrum. 
 The significance of the detection is quantified by calculating the probability for $A_L \leq 0$, $P(A_L \leq 0)$.
 As $A_L = 0$ is far out in the tail of the likelihood distribution of this parameter, we use high-temperature MCMC's chains to estimate $P(A_L \leq 0)$.
 Using SPT data only, we measure $P^{\{SPT\}}(A_L \leq 0) < \alensProbSpt$, the equivalent of a $\alensCdfSpt\, \sigma$ preference for $A_L > 0$ in a Gaussian distribution. 
 For SPT+\wseven{} we measure 
\begin{equation}
  P^{\{CMB\}}(A_L \leq 0) \, \leq \, \alensProbCmb \, ,
\end{equation}
 which corresponds to a $\alensCdfCmb\, \sigma$ detection of lensing in a Gaussian distribution.

Next, we report constraints on $A_L$.
 Using SPT+\wseven{} data we find
\begin{equation}
A_L = \alensCmb \, ,
\end{equation}
 where asymmetric $1\,\sigma$ (68.3\%) and $2\,\sigma$ (95.5\%) errors are shown.
 The observed lensing amplitude is consistent at $1\,\sigma$ with theoretical predictions in the \LCDM{} model.

\subsection{Mean Curvature of the Observable universe}
\label{sec:curvature}

\begin{figure*}
\begin{center}
    \includegraphics[angle=90,width=0.98\textwidth]{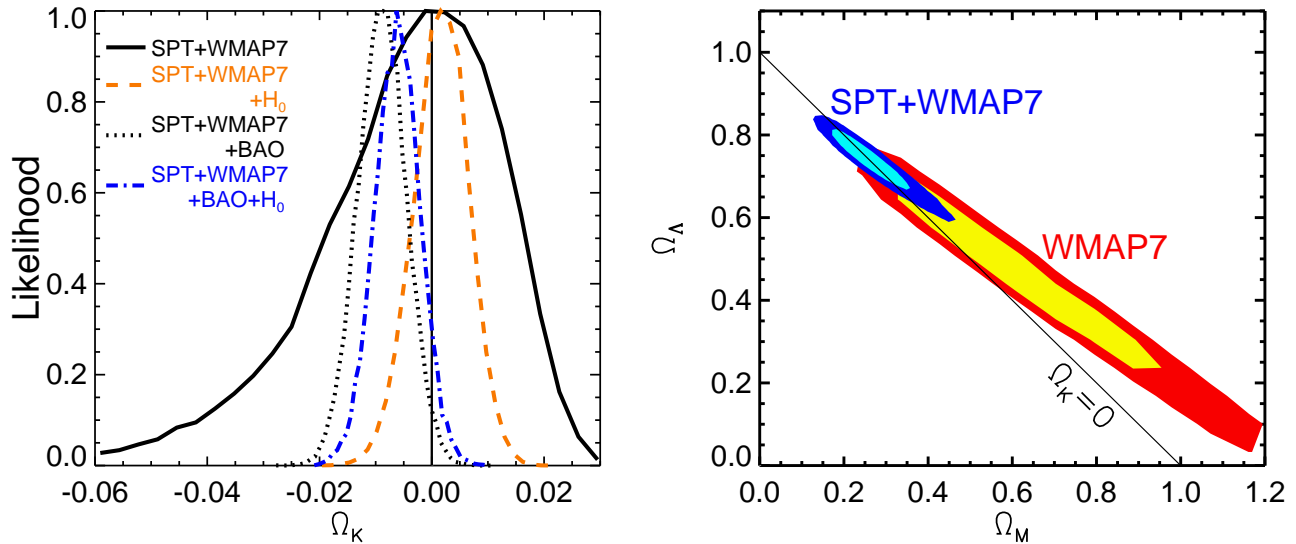}
\end{center}
\caption{
The low-redshift information imprinted on the CMB by gravitational lensing, along with the other information in the CMB anisotropy power spectrum, enables the placement of tight constraints on the mean curvature of the observable Universe.
The addition of low-redshift probes further tighten CMB-only constraints on the mean curvature. 
  \textbf{\textit{Left panel:}} The one-dimensional marginalized constraints on \omk{} from SPT+\wseven{} (\textbf{black solid line}), SPT+\textit{WMAP}7+\ho (\textbf{orange dashed line}), SPT+\textit{WMAP}7+BAO (\textbf{black dotted line}), and SPT+\textit{WMAP}7+BAO+\ho{} (\textbf{blue dot-dashed line}).
 The SPT+\wseven{} datasets measure the mean curvature of the observable universe to a precision of $\sim$1.5\%, while combining SPT,\wseven, and either \ho{} or BAO data reduces the uncertainty by a factor of $\sim$3. 
  \textbf{\textit{Right panel:}} The two-dimensional constraints on $\Omega_M$ and $\Omega_\Lambda$ from the SPT+\wseven{} data alone.
 The SPT+\wseven{} data rule out $\Omega_{\Lambda} =0$ at $\omlCdfCmb\,\sigma$.
}
\label{fig:omk}
\end{figure*}

The low-redshift information imprinted on the CMB by gravitational lensing, along with the other information in the CMB anisotropy power spectrum, enables the placement of tight constraints on the mean curvature of the observable Universe.
 The magnitude of the mean curvature today can be parameterized by $\omk \equiv -K/H_0^2$ where $\sqrt{1/|K|}$ is the length scale over which departures from Euclidean geometry become important.  
 Inflationary models generically predict $|\omk| \la 10^{-5}$ \citep[e.g.,][]{knox06}; thus a significant measurement of $\omk \neq 0$ would challenge our standard picture of the very early universe.
 A positive value for \omk ($K < 0$) could be obtained by the nucleation of a bubble of lower vacuum energy in a surrounding medium with higher vacuum energy followed by a short period of inflation \citep{bucher95}. 
 A determination that \omk{} is negative with high statistical significance would be very interesting; such a detection would be difficult to understand in the theoretical framework of inflation, challenge the string theory landscape picture, and rule out the de Sitter equilibrium cosmology of \citet{albrecht11}.

Absent lensing effects, one can leave the CMB power spectrum unchanged while simultaneously varying \omk{} and $\Omega_\Lambda$ in a way that keeps the distance to last scattering fixed \citep{bond97,zaldarriaga97c}.\footnote{There is an exception to this at very large scales due to the late ISW effect, but sample variance makes these changes unobservably small in the CMB power spectrum.}  
 Historically the CMB data placed very coarse constraints on \omk, with finer constraints only possible with the addition of other data sensitive to \omk{} and $\Omega_\Lambda$ such as $H_0$ measurements and determinations of $\Omega_m$ \cite[e.g.,][]{dodelson00} from, for example, the baryon fraction in clusters of galaxies \citep{white93a}.  

The sensitivity of the CMB to low-redshift information through gravitational lensing makes it possible to constrain the mean curvature of the observable universe, and thus the cosmological constant, using the CMB alone.
 The lensing amplitude is sensitive to the distance and growth of structure at intermediate redshifts ($0.5 \lesssim z \lesssim 4$).
 These observables are, in turn, sensitive to curvature, dark energy, and neutrino masses, as discussed in H12.  
 The recent detections of CMB lensing have measured an amplitude that is consistent with $\Omega_K$$\sim$$0$ and $\Omega_\Lambda$$\sim$$0.7$ \citep{das11b, das11, sherwin11, keisler11, vanengelen12}.  
 Simply put, the strength of CMB lensing in a universe with no dark energy and positive mean curvature would be much larger than that observed (see e.g., \citealt{sherwin11}).  

Using the SPT+\wseven{} bandpowers, we measure the mean curvature of the observable universe using only the CMB:

\begin{equation}
\label{eqn:omk_cmb}
 \omk = \omkCmb\, .
\end{equation} 
This result tightens curvature constraints over \wseven{} combined with low-redshift probes by $\sim$20\%.
 This constraint is consistent with zero mean curvature, and corresponds to a dark energy density of $\Omega_\Lambda=0.740^{+0.045}_{-0.054}$ and Hubble constant, $H_0=70.9^{+9.2}_{-8.0}$ (km s$^{-1}$ Mpc$^{-1}$).  
 This measurement rules out $\Omega_{\Lambda} =0$ at $\omlCdfCmb\,\sigma$ using the CMB alone.
 The right panel of Figure~\ref{fig:omk} shows the corresponding two-dimensional marginalized constraints on $\Omega_M$ and $\Omega_{\Lambda}$.
 We have confirmed that the strength of this constraint relies on the lensing signal; allowing $A_L$ to vary freely causes the curvature constraint to degrade dramatically.

CMB lensing enables an independent constraint on curvature, although the most powerful curvature constraints still come from combining CMB data with other low-redshift probes (e.g., \ho{}, BAO).
 The curvature constraint using CMB+\ho{} data is $\omk = \omkCmbHo$, while the constraint using CMB+BAO data is $\omk = \omkCmbBao$.  
 The tightest constraint on the mean curvature that we consider comes from combining the CMB, \ho{}, and BAO datasets:

\begin{equation}
\label{eqn:omk_cmb_h0_bao}
 \omk = \omkCmbHoBao\,.
\end{equation}

While the CMB+BAO constraint shows a $2.0\, \sigma$ preference for $\omk<0$, the significance of this preference decreases as more data are added. 
 The tightest constraint, coming from CMB+\ho{}+BAO, is consistent with zero mean curvature at $1.5\, \sigma$.
 These results are summarized in Figure~\ref{fig:omk}.  
 As discussed by H12, other extensions of \LCDM can also explain the data (e.g., allowing for both non-zero mean curvature and a dark energy equation of state $w \ne -1$), thus these constraints are significantly degraded when multiple extensions to \LCDM{} are simultaneously considered.


\subsection{Inflation}
\label{sec:inflation}

\begin{table*}
\begin{small}
\begin{center}
\caption{Constraints on \ns{} and $r$ from CMB and external datasets}
\begin{tabular}{l | c | c c c c}
\hline \hline
Model & Parameter & CMB & CMB+$H_0$ & CMB+BAO & CMB+$H_0$+BAO \\
 & & \footnotesize{(SPT+\wseven)} \\
\hline
\LCDM            & $n_s$              & \nsCmb            &    \nsCmbHo            &    \nsCmbBao           &    \nsCmbHoBao \\
\hline
\LCDMnospace+$r$ & $n_s$              & $0.969 \pm 0.011$ &    $0.9702 \pm 0.0097$ &    $0.9553 \pm 0.0084$ &    $0.9577 \pm 0.0084$ \\
                 & $r$ (95\% C.L.)      & $<\,0.18$ & $< 0.18$ & $< 0.11$  & $< 0.11$ \\
\hline
\hline
\end{tabular}
\label{tab:nsr}
\end{center}
\end{small}
\end{table*}

Cosmic inflation is an accelerated expansion in the early universe \citep{guth81,linde82,albrecht82} that generically leads to a universe with nearly zero mean curvature and a nearly scale-invariant spectrum of ``initial'' density perturbations \citep{mukhanov81,hawking82,starobinsky82,guth82,bardeen83} that evolved to produce the observed spectrum of CMB anisotropies.
 Models of inflation compatible with current data generally predict, over the range of observable scales, scalar and tensor perturbations well characterized by a power law in wavenumber $k$,

\begin{equation}
   \Delta^2_{R}(k) = \Delta^2_{R}(k_0) \left( \frac{k}{k_0}\right)^{{\displaystyle n_s}-1}
\end{equation}
\label{eqn:ns}
\begin{equation}
   \Delta^2_{h}(k) = \Delta^2_{h}(k_0) \left( \frac{k}{k_0}\right)^{{\displaystyle n_t}} \, .
\end{equation}
Here  $\Delta_R^2(k_0)$ is the amplitude of scalar (density) perturbations specified at pivot scale $k = k_0$, with scale dependence controlled by the index \ns{}, 
while $\Delta_h^2(k_0)$ is the amplitude of tensor (gravitational wave) perturbations specified at the same pivot scale, with scale dependence set by $n_t$.  
 The amplitude of the tensor perturbation spectrum is expressed in terms of the tensor-to-scalar ratio
\begin{equation}
\label{eqn:r}
 r =\left.\frac{\Delta_h^2(k)}{\Delta_R^2(k)} \right|_{k=0.002 {\rm\, Mpc}^{-1}}\, .
\end{equation}
 For single-field models in slow-roll inflation, $n_t$ and $r$ are related by a consistency equation  \citep{copeland93,kinney08}:
\begin{equation}
   n_t=-r/8 \, .
\end{equation}
The tensor and scalar perturbations predicted by such models of inflation can thus be characterized by the three parameters \ns{}, $\Delta^2_{R}(k_0)$,  and $r$.

In the following, we first consider constraints on \ns{} assuming $r=0$, and then on both $n_s$ and $r$.
 We then compare the constraints in the $n_s$-$r$ plane to predictions of inflationary models. 
 Constraints on the scale dependence of the spectral index (\nrun) are considered in the companion paper H12.

%
%
\subsubsection{Constraints on the Scalar Spectral Index}

\begin{figure*}
\begin{center}
     \includegraphics[width=0.98\textwidth]{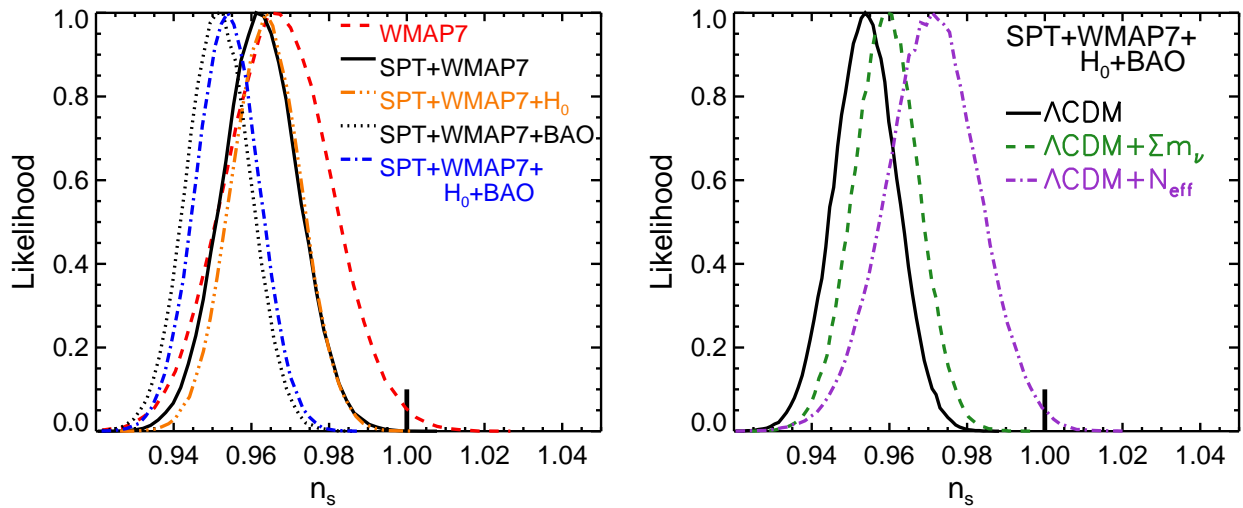}
\end{center}
\caption{The data strongly prefer departures from a scale invariant primordial power spectrum ($n_s < 1$), as predicted by inflation. 
 \textbf{\textit{Left panel:}} The marginalized one-dimensional constraints on \ns{} for the standard \LCDM model (with $r=0$) using several datasets.
 SPT data tightens the constraint on \ns{} relative to \wseven{} alone.
 Adding BAO data further tightens this constraint and leads to a preference for lower values of \ns{}, while adding \ho{} has little effect.
 \textbf{\textit{Right panel:}} The one-dimensional marginalized constraints on \ns{} from the SPT+\textit{WMAP}7+$H_0$+BAO dataset given three different models.  
 Plotted are \LCDM (\textbf{black solid  line}),  $\Lambda$CDM+\sumnu (\textbf{purple dashed  line}) as a typical case for extensions affecting the late-time universe, 
 and $\Lambda$CDM+\neff (\textbf{green dot-dashed  line}) as a typical case for extensions affecting the Silk damping scale.   
Of the extensions considered here, only those that affect the damping tail -- in this case by varying neutrino species -- causes noticeable movement towards $\ns = 1$.
 We note that in all cases the data robustly prefer a scale-dependent spectrum with $\ns < 1$.
}
\label{fig:ns}
\end{figure*}

Inflation is a nearly time-translation invariant state, however this invariance must be broken for inflation to eventually come to an end.
 The wavelength of perturbations depends solely on the time that they were produced, thus a time-translation invariant universe would produce scale-invariant perturbations ($\ns = 1$).\footnote{Scale invariance here means that the contribution to the rms density fluctuation from a logarithmic interval in $k$, at the time when $k = aH$, is independent of $k$.  Here $a(t)$ is the scale factor and $H\equiv \dot{a}/a$ is the Hubble parameter.}
 The prediction that inflation should be nearly, but not fully, time-translation invariant gives rise to the prediction that \ns{} should deviate slightly from unity \citep{kinney97a}.

Because of the special status of $n_s = 1$, and because we generally expect a departure from $n_s = 1$ for inflationary models, detecting this departure is of great interest.  
 K11 combined data from SPT and \wseven{} to measure a $3.0\,\sigma$ preference for $n_s<1$ in a \LCDM{} model, with $n_s~=~0.966\pm0.011$.  
 We show our constraints on \ns{} for the \LCDM{} model from several combined datasets in the left panel of Figure~\ref{fig:ns}.
 All datasets strongly prefer $\ns < 1$.

Using SPT+\wseven{} data, we find 
\begin{equation}
   \ns = \nsCmb \, .
\end{equation}
 For this dataset, we find $P(\ns > 1) = \nsProbCmb$, a $\nsCdfCmb\,\sigma$ departure from $n_s = 1$; $n_s < 1$ is heavily favored.

Including BAO data substantially shifts and tightens the constraints on \ns{}, as can be seen in Figure~\ref{fig:ns}.  
 The BAO distance measure $r_s/D_V$ depends on $\Omega_\Lambda$, breaking the partial degeneracy between $\Omega_\Lambda$ and \ns{} in the CMB power spectrum.
 The BAO preference for lower $\Omega_\Lambda$ pulls the central value of \ns{} down to $\ns = \nsCmbBao$.
 Using a high-temperature MCMC, we measure the probability for \ns{} to exceed one to be $\nsProbCmbBao$, corresponding to a $\nsCdfCmbBao\,\sigma$ detection of $n_s < 1$.

Including \ho{} data has a smaller effect on the \ns{} constraint than BAO, slightly disfavoring low-\ns{} values as seen in Figure~\ref{fig:ns}.
 The mechanism for the improvement is the same as for BAO, however, the CMB and \ho{} datasets individually prefer similar values of $\Omega_\Lambda$.
 Thus the two datasets tighten the \ns{} constraint around the CMB-only value, leading to $\ns = \nsCmbHo$.
 Using the combined CMB+\ho dataset, we measure the probability for \ns{} to exceed one to be \nsProbCmbHo, corresponding to a $\nsCdfCmbHo\,\sigma$ preference for $n_s < 1$.

As expected, combining CMB with both BAO and \ho{} data nudges the constraint on \ns{} up slightly from the CMB+BAO constraint to $\ns = \nsCmbHoBao$, thus weakening the preference for $n_s < 1$ slightly from $\nsCdfCmbBao$ to $\nsCdfCmbHoBao\,\sigma$.

In summary, regardless of which datasets we use, the data strongly prefer $\ns < 1$ in the \LCDM model.

The importance of detecting a departure from scale invariance leads us to review our modeling assumptions.  
 Specifically, are there extensions to the standard \LCDM model that could reconcile the data with a scale-invariant spectrum, $n_s=1$?
 We answer this question by calculating the \ns{} constraints from the CMB+$H_0$+BAO dataset for several physically motivated \LCDM model extensions.

We consider two classes of model extensions: those that can affect the slope of the CMB damping tail, and those that cannot.
 As a representative case of the first class of extensions, we consider \LCDMnospace+\neff, in which the number of relativistic species is allowed to vary. 
 As an example of the second class of extensions, we consider massive neutrinos \LCDMnospace+\sumnu (with \neff{} fixed at its fiducial value of 3.046).  
 These example extensions as well as several others are explored in considerable detail in H12.

Of the extensions considered, only models that can affect the slope of the damping tail significantly increase the likelihood of $\ns = 1$.
 The results of this test are displayed in the right panel of Figure~\ref{fig:ns}, where we show the marginalized constraints on \ns{} from the CMB+BAO+\ho{} dataset.
 Even the \neff extension does a poor job reconciling the data with a scale-invariant spectrum; the cumulative probability for $\ns > 1$ is \nsProbCmbHoBaoNeff, constraining \ns{} to be less than one at $\nsCdfCmbHoBaoNeff\,\sigma$. 
 The preferred value for \neff is far greater than the nominal value of 3.046 in the limited allowed parameter space.

We conclude that the data robustly prefer a scale-dependent spectrum with $n_s < 1$.

\subsubsection{Constraints on Tensor Perturbations}
\label{sec:ns_r}

\begin{figure}
\begin{center}
    \includegraphics[width=0.48\textwidth]{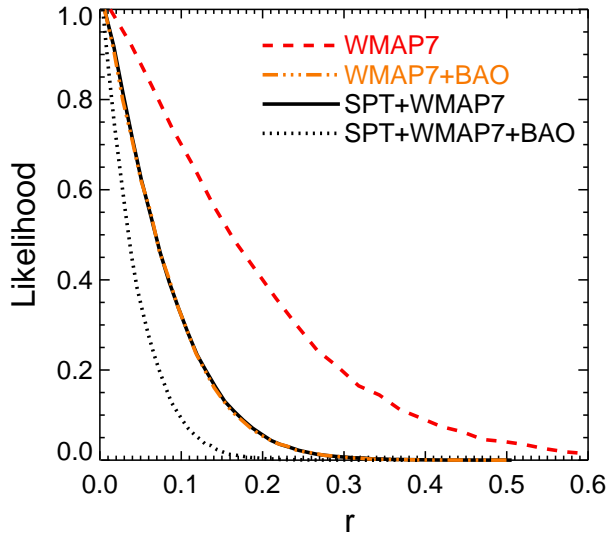}
\end{center}
\caption{This figure highlights the contributions of the SPT data to constraints on the tensor-to-scalar ratio, $r$. 
 We show four datasets: \wseven{} (\textbf{red dashed line}), \textit{WMAP}7+BAO (\textbf{orange dot-dashed line}), SPT+\wseven{} (\textbf{black solid line}), and SPT+\textit{WMAP}7+BAO (\textbf{black dotted line}). 
 Note that the \textit{WMAP}7+BAO and SPT+\wseven{} likelihood functions are nearly identical.
 SPT data tightens the $r$ constraint significantly, regardless of whether BAO data are included.
 While adding low-redshift $H_0$ measurements has minimal effect on the constraints on $r$ (not shown),  adding low-redshift information from BAO  tightens constraints on $r$ considerably.
 SPT+\wseven{} constrains $r < 0.18$ (95\% C.L.), while adding low-redshift BAO measurements tightens the constraint to $r < 0.11$ (95\% C.L.).}
\label{fig:r}
\end{figure}

The last inflationary parameter we consider is the tensor-to-scalar ratio $r$.
 Because $r$ is related to the energy scale of inflation,\footnote{The tensor-to-scalar ratio $r$ is proportional to the inflaton potential $V(\phi)$ and the energy scale of inflation is proportional to $V(\phi)^{1/4}$ \citep{kinney03, baumann08}.} 
 a detection of $r$ would provide an extremely interesting window onto the early universe.  
 We first consider the marginalized constraints on $r$ for the \LCDMnospace+$r$ model, shown in Table~\ref{tab:nsr} and Figure~\ref{fig:r}, then move on to a comparison with inflationary models in the $n_s$-$r$ plane in \S~\ref{sec:inflation_models} and Figure~\ref{fig:r_ns}.

One can think of the $r$ measurement in the following way.
 The CMB power spectrum is first measured at $\ell \ga 60$ (where tensor perturbations are negligible) to determine the \LCDM{} model parameters, thus determining the scalar contributions to the power spectrum.
 This scalar-only spectrum is then extrapolated to low $\ell$; any excess power observed is due to tensor perturbations.
 In this way, although the SPT data presented here do not directly measure power that could be from gravitational waves, by pinning down other model parameters, the extrapolation of the scalar power to large scales is more precise.  

Even with the parameters that determine the scalar power spectrum perfectly known, there would still be significant uncertainty in the value of $r$ due to cosmic variance.
 \citet{knox94} showed that for different realizations of a universe where $r=0$, 50\% of cosmic variance limited full sky temperature surveys will be able to place a limit of $r<0.1$ at 95\% confidence.
 As we will see, the results we present here approach that limit.  

\wseven{} data alone have been used to constrain $r < 0.36$ at 95\% confidence in the \LCDMnospace+$r$ model \citep{larson11}.
Prior to the current SPT analysis, the tightest published constraint on $r$ was reported by \citet{sanchez12}, who used the combination of datasets from BOSS-CMASS, \wseven, and K11 to constrain $r<0.16$ at 95\% confidence.

The bandpowers presented here lead to a significant reduction in the upper limit on $r$. 
 These measurements are summarized in Table~\ref{tab:nsr}.
 As shown in Figure~\ref{fig:r}, SPT bandpowers tighten the constraint on $r$ regardless of whether low-redshift information from BAO or \ho is included.
 Using the CMB datasets, we measure 
\begin{equation}
   r<0.18 \,\,(95\% \, \rm C.L.) \, .
\end{equation}
 This limit remains unchanged for the CMB+$H_0$ datasets, while from the CMB+BAO datasets we measure $r<0.11$ (95\% C.L.).
 As with constraints on \ns{}, the improvement with the addition of BAO comes from the improved BAO constraints on $\Omega_{\Lambda}$ breaking a partial three-way degeneracy between $\Omega_\Lambda$, \ns, and $r$.
 In contrast, adding in the $H_0$ measurement to CMB data does not significantly tighten the CMB-only result because the $H_0$ measurement sharpens up the $\Omega_\Lambda$ distribution around the values that allow for larger $r$.  
 It is worth noting that the SPT bandpowers tighten the constraint on $r$ even in the presence of BAO data; this can be seen in the tighter $r$ constraint from the \textit{WMAP}7+BAO to the SPT+\textit{WMAP}7+BAO datasets in Figure~\ref{fig:r}.

Using the combination of the CMB+BAO+$H_0$ datasets, we measure 
  \begin{equation}
   r<0.11 \,\,(95\% \, \rm C.L.) \, ;
 \end{equation}
 this limit is unchanged from the CMB+BAO constraint.
 The \ho{} and BAO data pull in opposite directions on the CMB-only data in the $r_s/D_V$ -- $H_0$ plane, however the statistical weight of the BAO measurements dominate the constraint in the combined dataset.

With the addition of low-redshift information from the BAO measurement, the constraints on tensor perturbations approach the theoretical limit of what can be achieved with temperature anisotropy alone.
 With the combination of these data, the cosmological parameters controlling the scalar perturbations are now sufficiently well known that they do not significantly degrade the limit on $r$.  
 This is evident in Figure~\ref{fig:r_ns}, where adding BAO removes most of the degeneracy between \ns{} and $r$.
 Further improvements now await precision $B$-mode CMB polarization observations \citep{seljak97,kamionkowski97}.
 The lowest upper limit on $r$ from $B$-modes is currently $r<0.7$ (95\% C.L.) from the BICEP experiment \citep{chiang10}.

\subsubsection{Implications for Models of Inflation}
\label{sec:inflation_models}

\begin{figure*}
\begin{center}
    \includegraphics[width=0.70\textwidth]{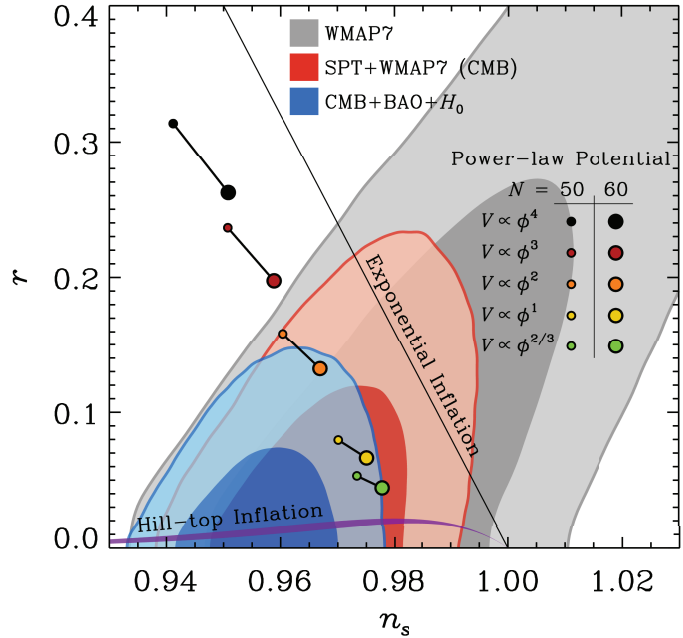}
\end{center}
\caption{ We compare the constraints on the \LCDMnospace+$r$ model with predictions from models of inflation in the $n_s-r$ plane.
 We show the two-dimensional constraints on $r$ and \ns{} as colored contours at the 68\% and 95\% confidence levels for three datasets: \wseven{} (\textbf{grey contours}), CMB (\textbf{red contours}), and CMB+$H_0$+BAO (\textbf{blue contours}).
 Adding the SPT bandpowers partially breaks the degeneracy between \ns{} and $r$ in the \wseven{} constraint, which can be seen clearly moving between the grey and red contours.
 Plotted over the constraint contours are predictions for several models of inflation.
 We restrict our comparison with model predictions to the simplest cases of slow-roll inflation due to a single scalar field as reviewed in \cite{baumann08}. \\
  \textbf{Solid black line}: The predictions of exponential inflation ($V(\phi) \propto \exp\left[\sqrt{16\pi\phi^2/(p\ M_{\rm Pl}^2)}\right]$) lie on this line.  In exponential inflation, increasing $p$ moves the prediction towards the Harrison-Zel'dovich-Peebles point $\ns = 1$, $r = 0$. \\
 \textbf{Black lines with colored circles}: The predictions of power-law potential inflation models ($V(\phi) \propto (\phi/\mu)^p$, $p>0$) for five different values of $p$ lie on the corresponding line. The predictions in the $r$ -- \ns{} plane are a function of $N$, where $N$ is expected to be in the range $N \in [50,60]$. \\
 \textbf{Purple region}: This region represents the upper limit on $r$ from large-field hill-top inflation models.  The width of the curve represents the uncertainty due to varying $N \in [50,60]$, but does does not take into account effects of higher order terms in the potential which may become important at the end of inflation.
}
\label{fig:r_ns}
\end{figure*}

We now turn to a comparison of model predictions with data constraints in the $n_s$-$r$ plane.
 This comparison is illustrated in Figure~\ref{fig:r_ns}, where we show the two-dimensional marginalized constraints from three combinations of data with predictions from simple models of inflation over-plotted.
 First, we note that the confidence contours for the CMB-only case in Fig.~\ref{fig:r_ns} show the expected positive correlation between \ns{} and $r$.
 Essentially, the suppression of large-scale power when increasing $n_s$ can be countered by adding extra large-scale power sourced by tensors. 
 The SPT data disfavor large values of $n_s$ (and hence $r$), significantly reducing the degeneracy between these two parameters. 
 With BAO data added to SPT+\wseven{}, the $\ns~-~r$ correlation nearly disappears. 
 Adding \ho{} data has little effect on constraints from the CMB or CMB+BAO datasets, removing only the smallest allowed values of \ns{} in both cases.
 As mentioned above, we are approaching the cosmic variance limit for the temperature anisotropy  on measuring $r$ -- at which point improved knowledge of the six \LCDM{} parameters no longer translates into better limits on $r$. 

We restrict the model comparisons  to the simplest cases of single-field, slow-roll inflation, as reviewed in \cite{baumann08}.  
 Models can be broadly characterized according to how much the inflaton field $\phi$ changes from the time perturbations on observably large scales were being produced until the end of inflation; this change in $\phi$ defined at $\Delta \phi$.  
 Models in which $\Delta \phi$ is larger than the Planck mass ($M_{\rm Pl}$) are classified as ``large-field'' models, while those in which $\Delta \phi < M_{\rm Pl}$ are classified as ``small-field'' models.
 The dividing line between the two cases  corresponds to $r = 0.01$.  

Here we look at large-field inflation models, considering several forms of the inflaton effective potential: 
 large-field power-law potential inflation models ($V(\phi) \propto (\phi/\mu)^p$, $p>0$), large-field hill-top inflation models ($V(\phi) \propto 1-(\phi/\mu)^2$), and exponential inflation models ($V(\phi) \propto \exp\left[\sqrt{16\pi\phi^2/(p\ M_{\rm Pl}^2)}\right]$).

Large-field power-law potential models have the fewest free parameters, and we discuss them first.
 Given a choice of $p$, these models have just one free parameter, and this parameter is highly constrained by the requirement of reproducing the well-known amplitude of the scalar perturbation spectrum.  
 Thus, these models make fairly localized predictions in the $\ns~-~r$ plane.  
 The uncertainty in these predictions is dominated by the details of the end of inflation which are not specified by $V(\phi)$ but, instead, depend on the coupling of the inflaton field $\phi$ to other fields.
 This uncertainty can be captured by the parameter $N$ where $e^N$ gives the increase in the scale factor between the time when the observable scale leaves the horizon and the end of inflation.\footnote{A note of clarification: $N$ says nothing about the total increase in the scale factor between the beginning and end of inflation, which is expected to be much larger.} 
 Assuming a standard slow-roll inflation scenario,\footnote{The assumption here is that inflation stops by the end of slow roll and is followed by the field oscillating in an approximately quadratic potential near the minimum.  The universe eventually reheats to a density greater than that during BBN, and then the standard thermal history ensues.} 
 $N$ is expected to lie in the range 50 to 60 \citep{liddle03}. 
 The spread in values is dominated by uncertainty in how much the energy density drops between the end of inflation and reheating, though this range can be extended in either direction by modifications to the standard thermal history.
 In Figure~\ref{fig:r_ns} we consider several large-field power-law potential models, each with a different value of $p$, and indicate the predictions of each model as $N$ varies between 50 and 60.

The $\lambda \phi^4$ ($p=4$) and $\lambda \phi^3$ ($p=3$) models were ruled out by earlier data \citep{komatsu09, dunkley11}, given the expected range of $N$.
 These models possibly could have been saved with a non-standard post-inflation thermal history (designed to make $N$ very large, thus moving the prediction towards $r=0$ and \ns=1 in Figure~\ref{fig:r_ns}), but such a maneuver no longer works.
 The CMB+BAO dataset excludes the $\lambda \phi^4$ model with greater than 95\% confidence for all values of $N$, while the $\phi^3$ ($p=3$) model is excluded with greater than 95\% confidence by the CMB or CMB+$H_0$ datasets given the expected range of $N$, and excluded with greater than 95\% confidence by the CMB+BAO datasets regardless of $N$.

While the $m^2\phi^2$ ($p=2$) model was consistent with previous constraints \citep{keisler11, sanchez12}, the current combinations of CMB with BAO data place a tight upper limit on $r$ and disfavor the $m^2 \phi^2$ model, which produces predictions that fall at the edge of the 95\% confidence contour.
 This model is allowed by CMB or CMB+\ho.
 Models with smaller values of $p$ are consistent with the data, as shown in Figure~\ref{fig:r_ns}.

The exponential models lead to \ns{} and $r$ predictions that are independent of scale and therefore independent of $N$.  
 The predictions, as $p$ varies, form a line in the $\ns~-~r$ plane.  
 This whole class of models is allowed at 95\% confidence for a range of $p$ by the CMB+\ho{} data, and is excluded ($> 95\%$ C.L.) by the CMB+BAO and CMB+$H_0$+BAO data.  

The potential in hill-top models has the shape of symmetry-breaking potentials that drive $\phi$ away from the origin.  
 The generic form of the hill-top potential is $V(\phi) \propto 1-(\phi/\mu)^p$.
Such models, for a fixed $p$, have three free parameters: the proportionality constant, $\mu$, and $\phi_{end}$.
The first two are found in the potential.
The third is needed because the potential does not naturally lead to an end to inflation; without making $\phi_{end}$ explicit, the end of inflation depends either on the details of unspecified higher order terms in the potential or on external physics. 
The proportionality constant is set by the amplitude of scalar fluctuations $\Delta_R^2$, while $\mu$, for fixed $\phi_{end}$, is constrained by  \ns{} and $r$.  
In turn, we take $\phi_{end}$ to be constrained above by $\mu$, of order the vacuum expectation value of the field.

For $p \le 2$, hill-top models can have the behavior of large-field models for the range of \ns{} allowed by the data.  The behavior of the $p =2$ case in the large-field ($\phi_{end} = \mu$) limit is shown as a purple region in Figure~\ref{fig:r_ns}, and is consistent with the data.

Finally, we consider small-field inflation models.
 Examples of small-field potentials include hill-top potentials with $p > 2$ (for some values of $\mu$), the Coleman-Weinberg potential \citep{coleman73} (with suitably adjusted parameters), and warped D-brane inflation \citep{kachru03}.  
 Because small-field models predict $r \le 0.01$, the SPT data have no constraining power on these models through limits on $r$.
 All of these models are consistent with the data, so long as they are consistent with the limits on \ns.

Constraints on the scale dependence of the spectral index (\nrun) are considered in the companion paper H12.

\section{Conclusion}
\label{sec:conclusion}

We present a measurement of the CMB temperature power spectrum from 2540 deg$^2$ of sky observed with the South Pole Telescope (SPT).
 These are the first CMB power spectrum results reported for the full SPT-SZ survey, 
which encompasses three times the area used in previous SPT power spectrum analyses (K11, \cite{reichardt13})
 The bandpowers cover the third to ninth acoustic peaks ($650 < \ell < 3000$) with sample-variance-limited precision at $\ell < 2900$.
 This measurement represents a significant advance over previous measurements of the damping tail by ACBAR \citep{reichardt09a}, QUaD \citep{brown09, friedman09}, ACT \citep{das11}, and SPT \citep{keisler11}.

We find the SPT bandpowers are well fit by a spatially flat \LCDM{} cosmology with gravitational lensing by large-scale structure.
 We use this SPT measurement to extend the dynamic range probed by the \wmap power spectrum, thus tightening parameter constraints in the six-parameter \LCDM model. 
 With the exception of the optical depth $\tau$ which is constrained by the large-scale polarization data from \wseven, adding the full survey SPT bandpowers significantly improves measurements of all \LCDM parameters.
 Most notably, the measurement of the angular sound horizon, $\theta_s$, tightens by a factor of 2.7 due to the number of acoustic peaks detected at high signal-to-noise.
 Uncertainties on the other four parameters are reduced by a factor of $\sim 1.4$.
 The combination of SPT and \wseven{} data is used to constrain $\ns < 1$ at $\nsCdfCmb\,\sigma$.

We examine constraints on three extensions to the \LCDM{} model. 
 We first allow for a rescaling of the gravitational lensing potential by a parameter $A_L$.
 Using CMB data, we rule out the no-lensing hypothesis ($A_L \leq 0$) at $\alensCdfCmb\,\sigma$, the most significant detection to date using the CMB alone, and measure a lensing amplitude, $A_L = \alensCmb$ (68\% and 95\% C.L.), consistent with the \LCDM{} expectations.
 We expect the lensing detection significance to triple in a future analysis of the full SPT-SZ survey using an optimized four-point lensing estimator, similar to the one applied to one fifth of the survey by \citet{vanengelen12}.

Second, we allow non-zero mean curvature of the observable universe.
 The low-redshift information encoded in the CMB by gravitational lensing helps to improve constraints on the mean curvature of the observable universe for the \LCDMnospace+\omk{} model.
 Using the CMB alone, we measure $\omk=\omkCmb$, which is consistent with a flat universe.
 Models without dark energy are ruled out at $\omlCdfCmb\,\sigma$.

Finally, we look at constraints on the amplitude of tensor perturbations.
 The combination of SPT+WMAP7 is used to constrain the tensor-to-scalar ratio to be $r < 0.18$ with 95\% confidence.

Adding low-redshift probes of \ho{} and BAO further tightens these constraints.
 Combining the CMB and \ho{} datasets mildly tightens our parameter constraints, and is fully consistent with the CMB constraints.
 Combining the CMB and BAO datasets, on the other hand, leads to significant improvements in our parameter constraints.
 Combining all three datasets produces constraints that lie close to the CMB+BAO constraints; 
 the combination of CMB+$H_0$+BAO is used to constrain $\ns < 1$ at $\nsCdfCmbHoBao\,\sigma$ in the \LCDM{} model, measure $\omk~=~\omkCmbHoBao$ in the \LCDMnospace+\omk{} model, and constrain $r < 0.11$ at 95\% confidence in the \LCDMnospace+$r$ model.
 This constraint on $r$ approaches the theoretical limit of how well tensor perturbations can be constrained from CMB temperature anisotropy, $r<0.1$ (95\% C.L.).
 We compare these constraints on $n_s$ and $r$ to the predictions of single-field inflation models and exclude several models with greater than 95\% confidence.

There is some tension between the six datasets included in the CMB+BAO+\ho{} combination for a \LCDM{} cosmology. 
 However, we assume the uncertainties reported for each of the datasets are correct and combine them to produce many of the results presented here. 
 We refer the reader to H12 for a more detailed discussion of the consistency of the datasets.

In this paper, we have focused on the amplitude and shape of the primordial power spectrum of scalar and tensor perturbations, as well as the effects of gravitational lensing and curvature.   
 Further cosmological implications of the bandpowers from the full SPT-SZ survey, including constraints on the neutrino masses, the dark energy equation of state, the primordial Helium abundance, and the effective number of neutrino species are  explored in the companion paper H12.

The first CMB temperature power spectrum from the {\it Planck} satellite is expected to be released in 2013.
 We expect that, given the small beam of the SPT relative to {\it Planck}, the data presented here should remain the most precise measurement of small angular scale anisotropy for $2200 \lesssim \ell <  3000$.
 The cosmological constraints presented here will tighten significantly with the with the {\it Planck} power spectra.  
 Further improvement will come with the addition of polarization information from upcoming experiments including {\it Planck}, ACT-Pol, Polarbear, and SPT-Pol, the new polarization-sensitive camera on the SPT.

\begin{acknowledgments}
We thank Scott Dodelson, John Peacock, David Baumann, and Antonio Riotto for useful conversations.
The SPT is supported by the National Science Foundation through grant ANT-0638937, with partial support provided by NSF grant PHY-1125897, the Kavli Foundation, and the Gordon and Betty Moore Foundation.
The McGill group acknowledges funding from the National Sciences and Engineering Research Council of Canada, Canada Research Chairs program, and the Canadian Institute for Advanced Research. 
Work at Harvard is supported by grant AST-1009012.
R. Keisler acknowledges support from NASA Hubble Fellowship grant HF-51275.01,
B.A.\ Benson a KICP Fellowship,
M.\ Dobbs an Alfred P. Sloan Research Fellowship,
O.\ Zahn a BCCP fellowship,
M. Millea and L. Knox a NSF grant 0709498.  
This research used resources of the National Energy Research Scientific Computing Center, which is supported by the Office of Science of the U.S. Department of Energy under Contract No. DE-AC02-05CH11231, and the resources of the University of Chicago Computing Cooperative (UC3), supported in part by the Open Science Grid, NSF grant NSF PHY 1148698.
Some of the results in this paper have been derived using the HEALPix \citep{gorski05} package. 
We acknowledge the use of the Legacy Archive for Microwave Background Data Analysis (LAMBDA). 
Support for LAMBDA is provided by the NASA Office of Space Science.
\end{acknowledgments}

{\it Facilities:}
\facility{South Pole Telescope}

\bibliography{../../BIBTEX/spt}

\appendix

\section{Dependence of SPT constraints on multipole range}
\label{sec:lrange}
\begin{figure*}
\begin{center}
    \includegraphics[angle=90,width=0.98\textwidth]{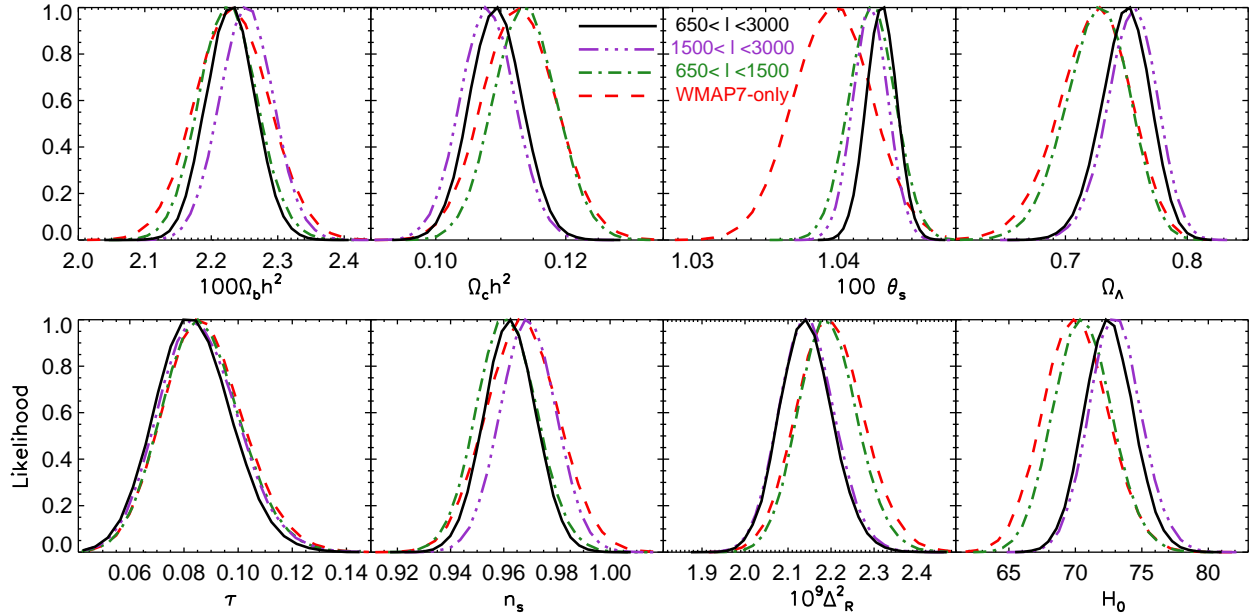}
\end{center}
\caption{
Same \LCDM-model parameters as Figure~\ref{fig:like1d_s12tau_w7}, except showing the effect of including the SPT bandpowers over sub-sets of the $\ell$-range.
The constraints for \wseven{}-only are shown (\textbf{red dashed lines}).
Using the SPT+\wseven{} dataset, the constraints are shown from the full $\ell$-range of SPT data (\textbf{black solid lines}), from the high-$\ell$ range $1500 < \ell_{SPT} < 3000$ (\textbf{purple dot-dot-dashed lines}), and from the low-$\ell$ range $650 < \ell_{SPT} < 1500$ (\textbf{green dot-dashed lines}).
}
\label{fig:like1d_lrange}
\end{figure*}

As discussed in \S~\ref{subsec:w7+s12} and shown in Figure~\ref{fig:like1d_s12tau_w7}, the \LCDM{} constraints tighten and shift from the \wseven{} to the SPT+\wseven{} dataset.
 We explore how different $\ell$-ranges of the SPT data drive these changes in Figure~\ref{fig:like1d_lrange}.
The shifts in $\Omega_c h^2$ and $\Omega_{\Lambda}$ (or $H_0$) are driven primarily by the bandpowers at higher multipoles, above $\ell=1500$.
 As discussed in H12, this preference is largely driven by the sensitivity of the SPT data to gravitational lensing in the high-$\ell$ acoustic peaks.
 On the other hand, \ns{} is largely constrained by the lower multipoles; the bandpowers at higher multipoles prefer slightly higher values of \ns.
 The SPT data prefer higher values of $\theta_s$ than the \wseven{} data; each of the sub-ranges pull $\theta_s$ above the WMAP value, but only the constraining power of the full dataset pulls $\theta_s$ up to the SPT+\wseven{} value.

\end{document}